\documentclass[twocolumn]{aastex62}
\usepackage{color} 
\usepackage{lipsum}
\usepackage{mathtools}
\usepackage{placeins}
\usepackage{multirow}
\usepackage{float}
\usepackage{amsmath}
\usepackage{physics}
\usepackage{silence}
\usepackage{bm}
\usepackage{multirow, booktabs}
\usepackage{rotating}

\shorttitle{The First 3$\pi$ 3D Map of ISM Dust Temperature}
\shortauthors{Zelko, Finkbeiner, Lee, Green}

\begin{document}
\title{The First 3$\pi$ 3D Map of Interstellar Dust Temperature}

\author{Ioana A. Zelko}
\footnote{Corresponding author: ioana.zelko@gmail.com}
\affiliation{Harvard-Smithsonian Center for Astrophysics, 
60 Garden Street,
Cambridge, MA  02138, USA}
\affiliation{Department of Physics and Astronomy, University of California-Los Angeles,
475 Portola Plaza, Los Angeles, CA 90095}
\affiliation{Canadian Institute for Theoretical Astrophysics, University
of Toronto, Toronto, ON, Canada M5S 3H8}
\author{Douglas P. Finkbeiner} 
\affiliation{Harvard-Smithsonian Center for Astrophysics,
60 Garden Street,
Cambridge, MA 02138, USA}
\affiliation{Department of Physics, Harvard University,
17 Oxford Street,
Cambridge, MA 02138, USA}
\author{Albert Lee}
\affiliation{Institute for Disease Modeling,
500 5th Ave N,
Seattle, WA 98109}
\author{Gregory Green}
\affiliation{Max Planck Institute for Astronomy,
Königstuhl 17, D-69117 
Heidelberg, Germany}

\begin{abstract}
We present the first large-scale 3D map of interstellar dust temperature. We build upon existing 3D reddening maps derived from starlight absorption (Bayestar19), covering 3/4 of the sky. Starting with the column density for each of 500 million 3D voxels, we propose a temperature and emissivity power-law slope ($\beta$) for each of them, and integrate along the line of sight to synthesize an emission map in five frequency bands observed by \emph{Planck} and \emph{IRAS}.  The reconstructed emission map is constrained to match observations on a $10'$ scale, and does so with good fidelity. We produce 3D temperature maps at  resolutions of $110', 55', $and $27'$. We assess performance on Cepheus, a dust cloud with two distinct components along the line of sight, and find distinct temperatures for the two components. We thus show that this methodology has enough precision to constrain clouds with different temperature along the line of sight up to $1-\sigma$ error. This would be an important result for dust frequency decorrelation foreground analysis for cosmic microwave background experiments, which would be impacted by a line-of-sight with varying temperature and magnetic field components. In addition to $T$ and $\beta$, we constrain the conversion factor between emission optical depth and reddening. This conversion factor is assumed to be constant in commonly used emission-based reddening maps. However, this work shows a factor of two variation that may prove significant for some applications. 

\end{abstract}


\section{Introduction}

Researchers have long used dust emission to map dust in 2D \citep{Schlegel1998,PlanckCollaboration2014,Collaboration2016a}, and in recent years starlight reddening has been used to map dust in 3D \citep{Green2015, Green2018, Green2019,Leike2019}. However, the technique of combining 3D information from reddening with emission data is still in incipient stage. 

Dust is primarily made of silicate and carbonaceous grains, with varying size distribution and composition across the sky \citep{Mathis1977, Greenberg1978, Cardelli1989, Desert1990,  Dwek1997, Li1997, Li2001, Weingartner2001a, Zubko2004, Draine2011, Jones2013, Wang2015}.  The dust is heated by the interstellar radiation field (ISRF), and emits in the far infrared (FIR). Absorption of radiation in the ultraviolet \citep{Savage1979, Fitzpatrick1999, Cardelli1989} and emission of FIR \citep{Reach1995,Finkbeiner1999, Bennett2003,PlanckCollaboration2014, Collaboration2016a} are the two main tracers of dust, and have both been used in numerous other studies. But dust absorption and emission do not trace each other, as composition, size, and ISRF vary \citep{Schlafly2016, Zelko2020}. Thus, mapping these properties of dust is critical.

\begin{figure*}[t!]
\begin{center}
\begin{tabular}{cccc}
    \hspace{-5.20mm}
	\includegraphics[width=0.5\textwidth]{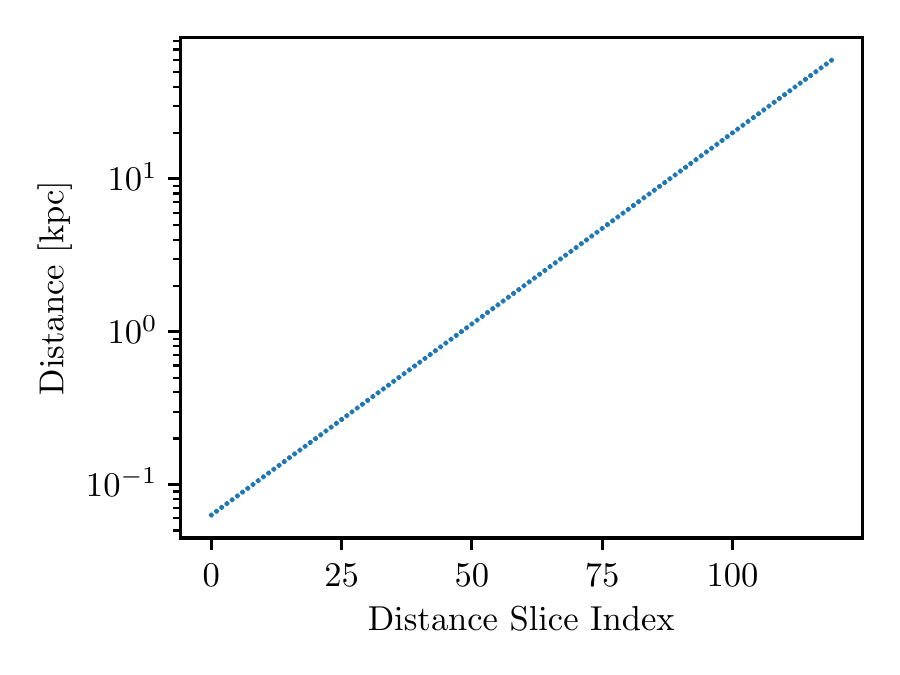} &
    \hspace{-4.70mm}
	\includegraphics[width=0.5\textwidth]{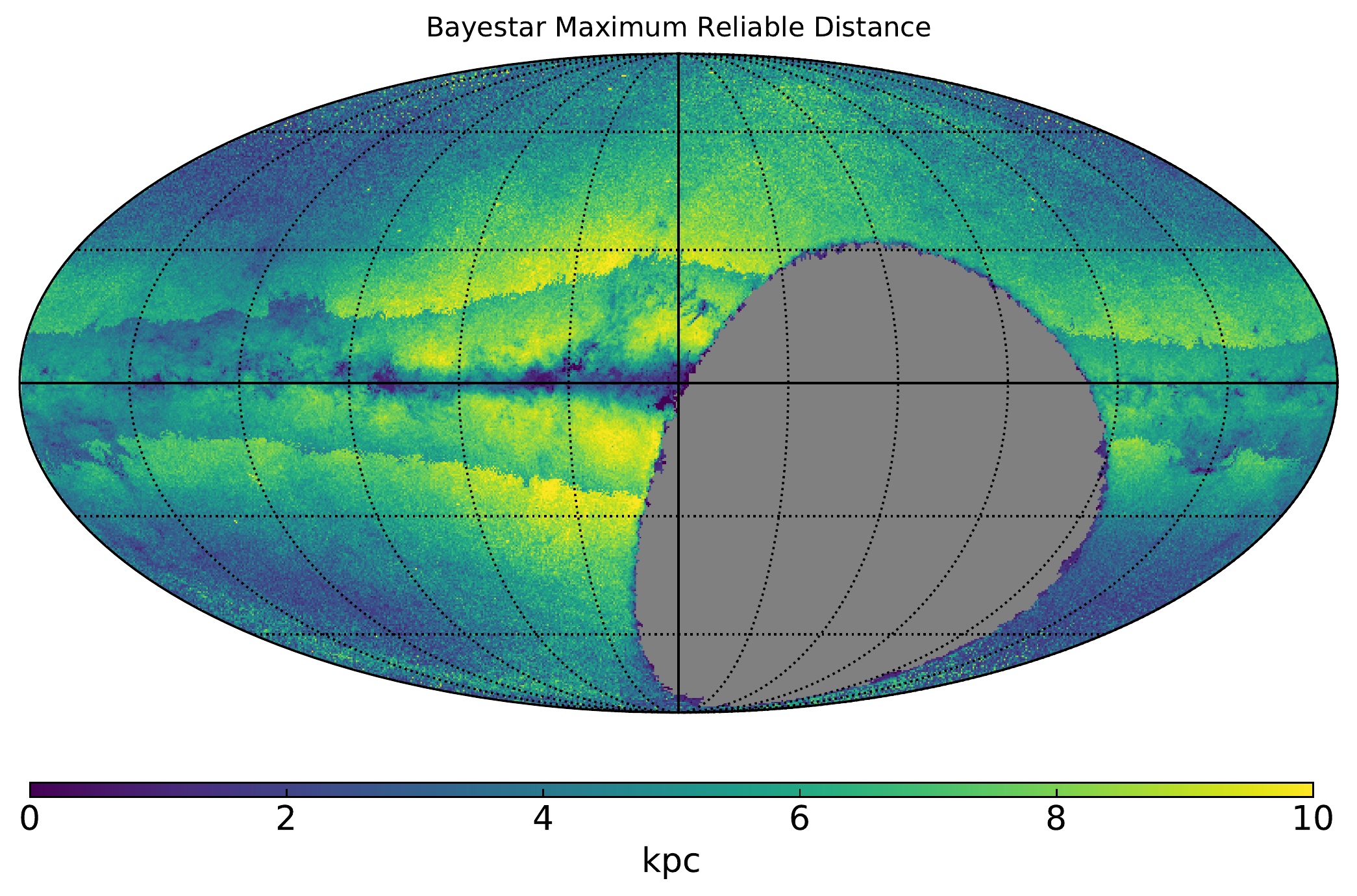}
\end{tabular}
\caption[Bayestar Distance Slices]{Bayestar \citep{Green2019} distance slices. The 3D map is sampled at 120 distance slices, linearly spaced in distance modulus. However, the number of bins that provide non-zero reddening varies greatly based on the direction in the sky, so the choice of the distance cuts in the 3D temperature map needs to be customized based on application and direction. The maximum value attained by the maximum reliable distance is 20.3 pc. The crispy boundaries show the limit where the resolution of the map varies across the sky, between Nside 256 and Nside 1024. \label{fig:bayestar_distances}}
\end{center}
\end{figure*}

\begin{figure*}[t!]
	\centering
	\begin{tabular}{cccc}
		\hspace{-5.20mm}
		\includegraphics[width=0.5\textwidth]{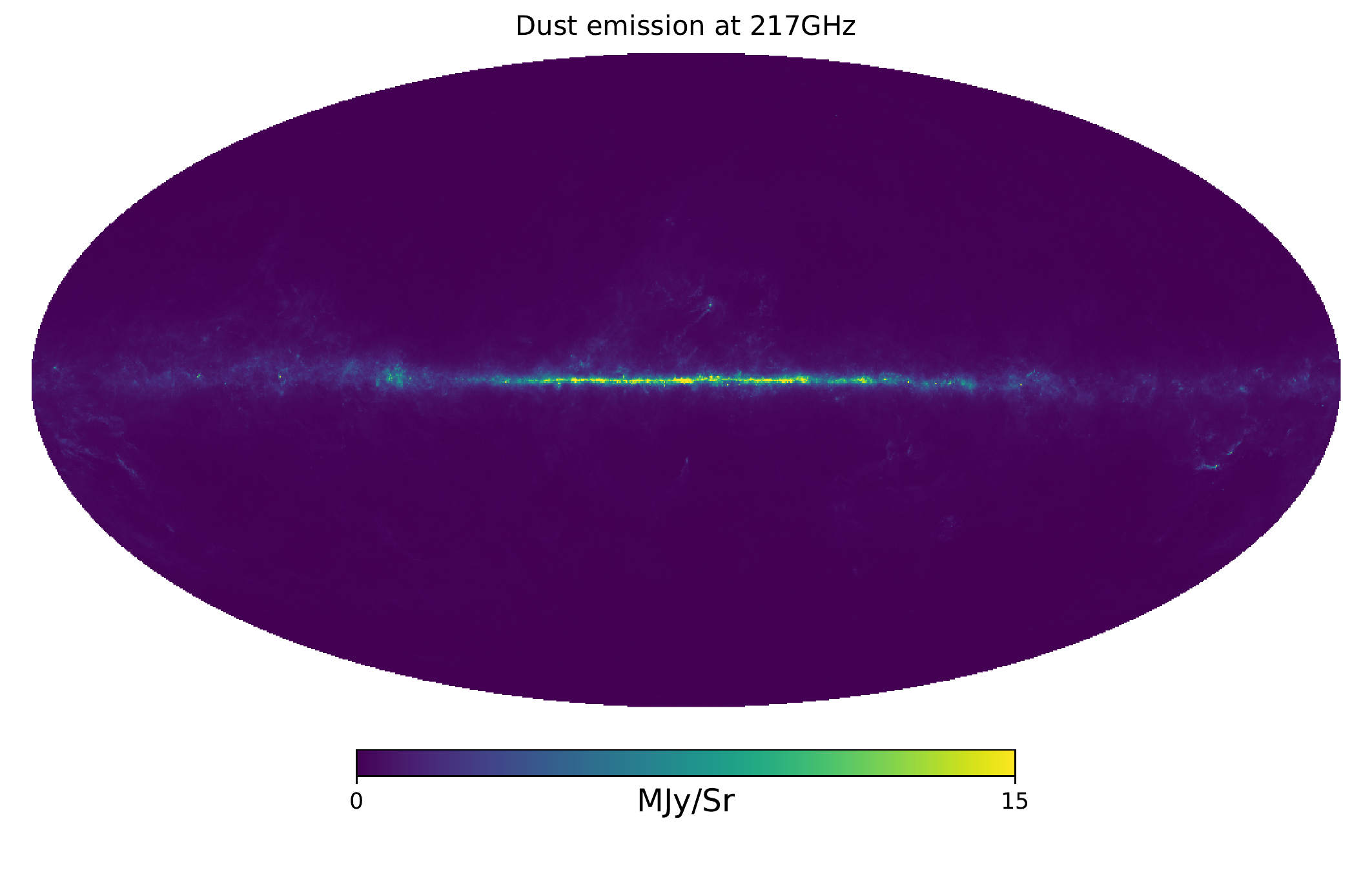} &
		\hspace{-4.70mm}	
		\includegraphics[width=0.5\textwidth]{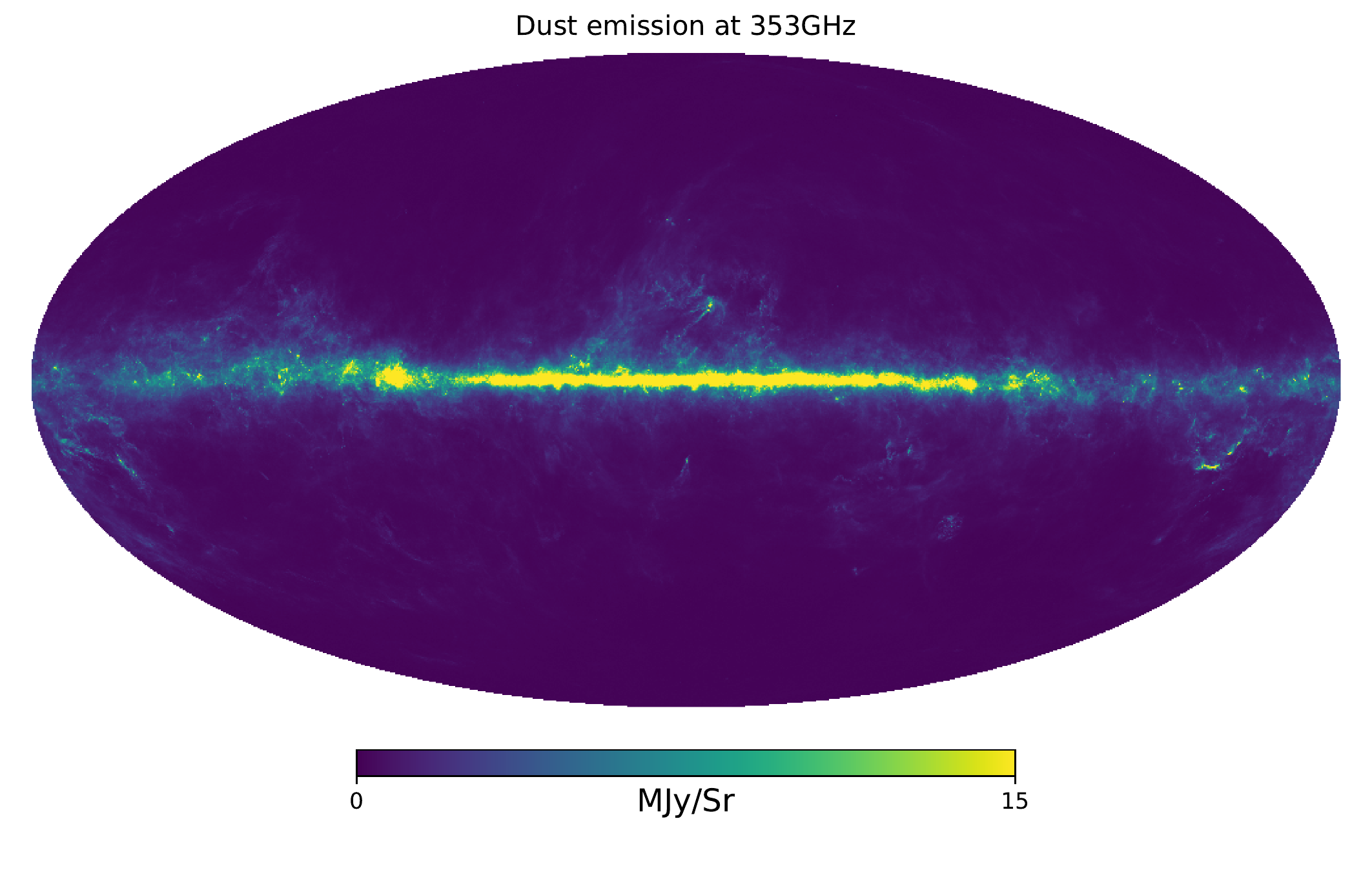} &
		\vspace{-2.00mm}
		
	\end{tabular}
	
	\vspace{-2.00mm}
	
	\begin{tabular}{cccc}
		\hspace{-5.20mm}
		\includegraphics[width=0.5\textwidth]{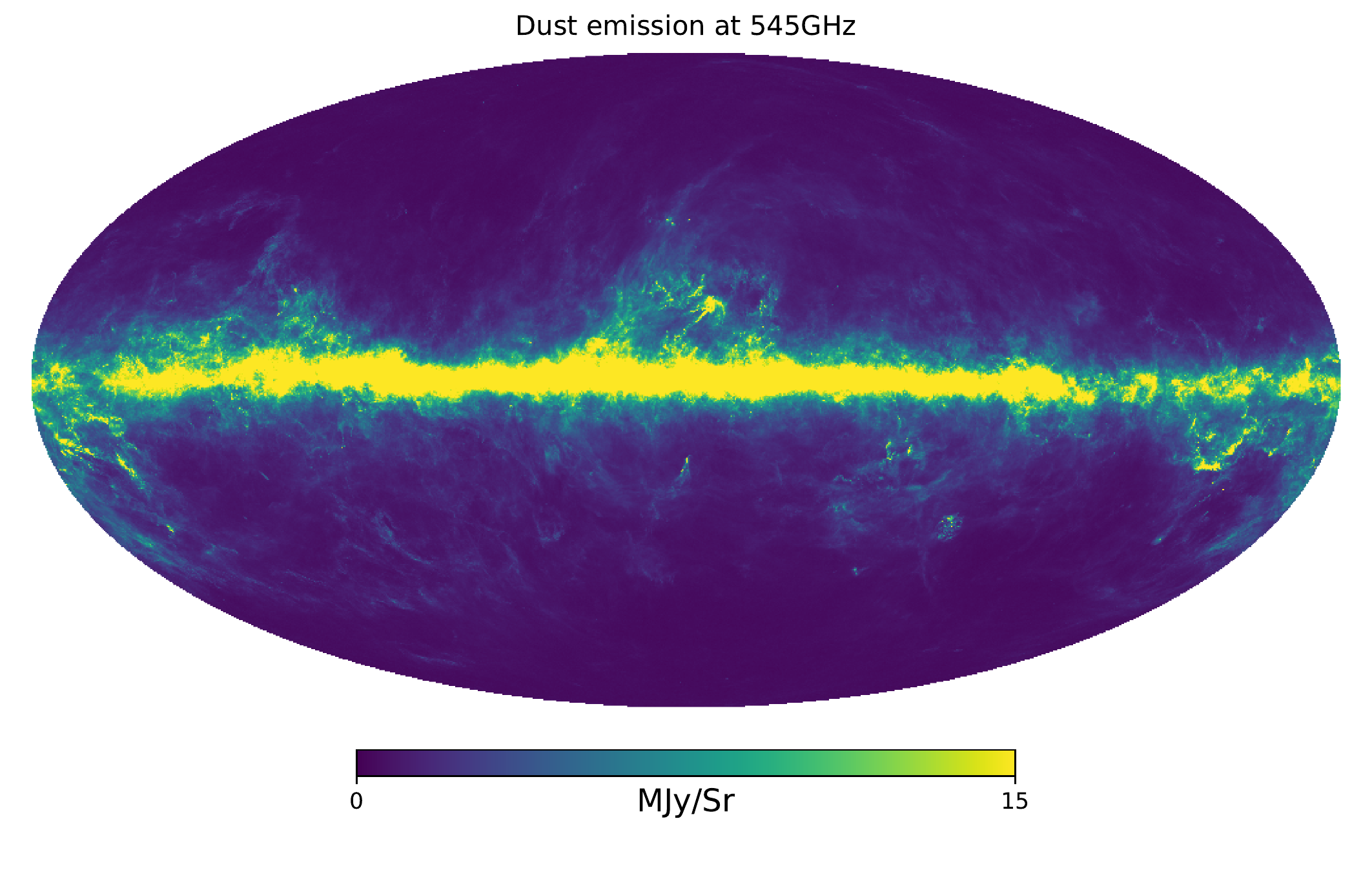} &
		\hspace{-4.70mm}		
		\includegraphics[width=0.5\textwidth]{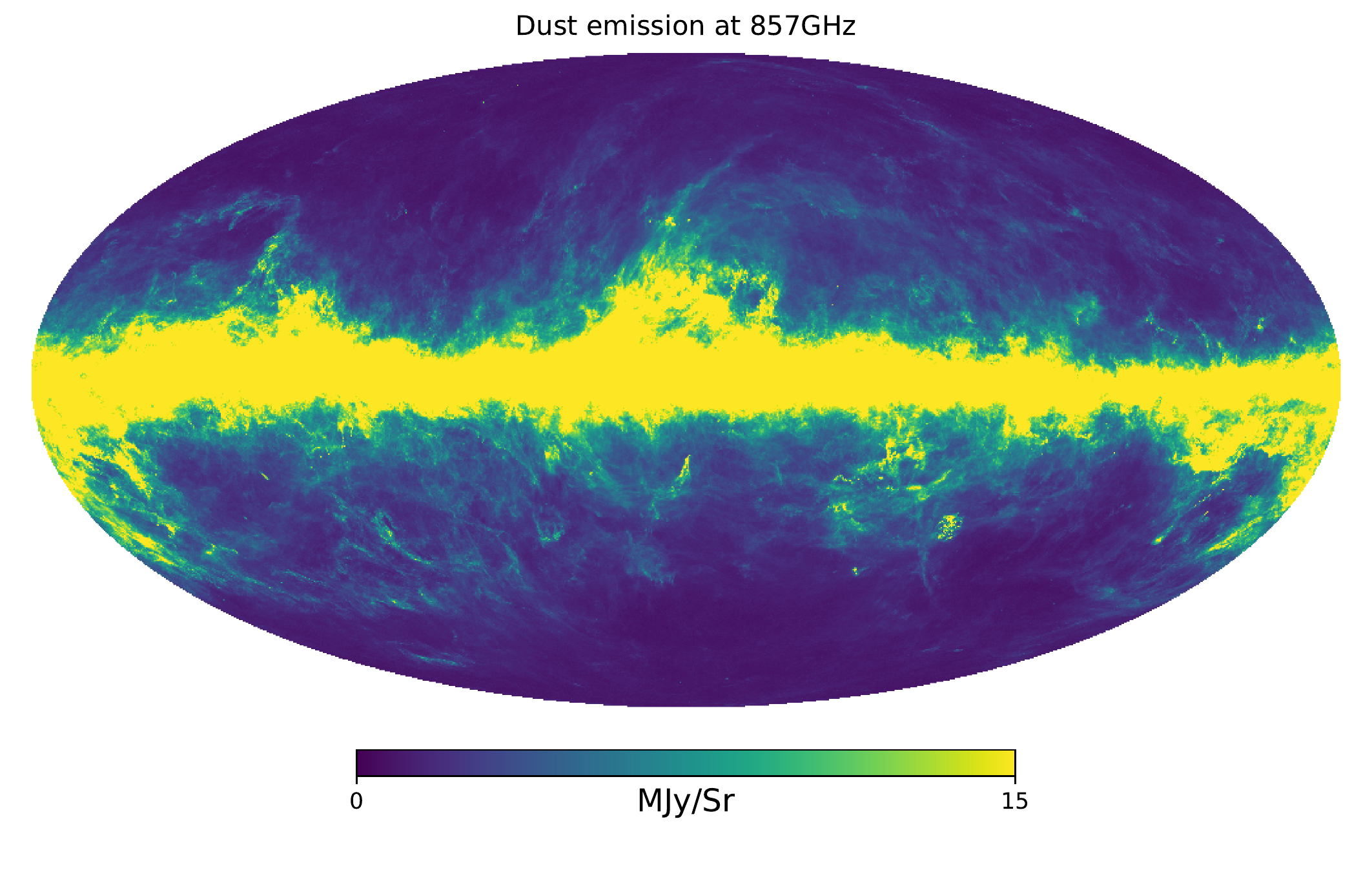} &

		\vspace{-2.00mm}
		
	\end{tabular}
	\vspace{-2.00mm}

    \hspace{-10.00mm}
	\includegraphics[width=0.5\textwidth]{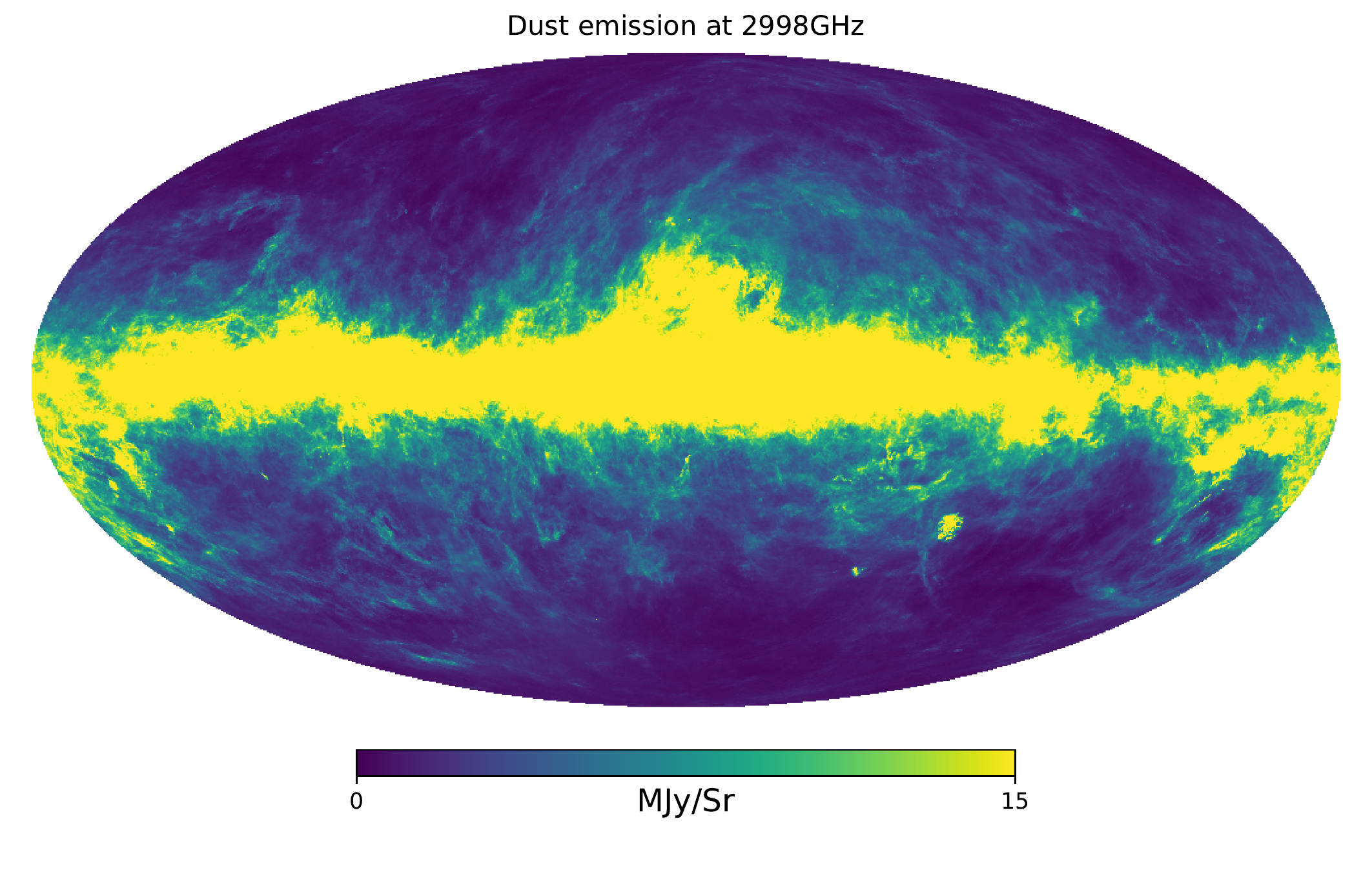}
	\caption[Sky Dust Emission Maps]{The sky-wide dust emission maps. The are obtained from the 217, 353, 545, 857GHz emission maps from the \emph{Planck}  satellite  and the 2998GHz \emph{IRAS/DIRBE} map, after subtracting the contribution of the CMB. The other foregrounds are subdominant to dust at these frequencies, with the exception of the CIB. This analysis thus applies only to the lines of sight where the contribution from the ISM dust is big enough to dominate the CIB.}
	
	\label{fig:dust_3D_emissison_data}
\end{figure*}

 Having a better understanding of the 3D dust can help us characterize the 3D structure of the magnetic field, and compute the six-dimensional phase-space density of the interstellar radiation field (the amount of light in every 3D voxel, in every direction, at every energy). Obtaining this data would allow us to see the Galaxy from any vantage point, at any angle, in any color. This 6-D radiation field is an input to anisotropic inverse-Compton scattering calculations done in gamma-ray astronomy, particularly important for dark matter annihilation searches. In addition, the 3D dust analysis in the galactic plane can provide us with critical information about correlation in dust properties, which can be used to inform the analysis at higher latitudes for cosmic microwave background(CMB) experiments, such as BICEP\citep{BICEPCollaboration2014,BICEPCollaboration2015}, CMB-S4 \citep{Abazajian2016}, PIXIE \citep{Kogut2006}, PRISM \citep{Andre2014}, CORE \citep{Finelli2018},LiteBIRD \citep{Matsumura2014} and other experiments. It is thus critical that we improve our characterisation of the dust, both for improving our knowledge of the ISM, and for cosmology.

In this work we present map with $3\pi$ sky coverage of the temperature of the galactic dust in 3D. To create it, we constrain the integrated dust emission along the line-of-sight (LOS) using full-sky emission observations. The challenge lies in disentangling how much emission comes from each distance along the LOS. Existing 3D reddening maps can help inform the distribution of dust along the LOS. However, density is not the only element that controls emission, and variation of dust temperature from cold to hot creates a degenerate problem. Previous work \cite{Martinez-Solaeche2018} has tried breaking the degeneracy by making assumptions about the shape of the magnetic field, and comparing to polarization data. We take a different approach. We assume that the dust emission parameters $T$ and $\beta$ (see Eq. \ref{eq:MBB_ISM_dust}) vary slowly with angle (for our "superpixel" scale corresponding to Nsides 32, 64, or 128). This allows us to use the much higher Nside 1024 resolution maps to identify the specific emissivity contribution coming from each volume element along the line of sight. This procedure, intuitively, separates structures/patterns in the density of the 3D dust maps along the LOS and breaks the emissivity degeneracy, allowing us put error bars on the temperature measurements. For now, this is a safer assumption compared to the degree of variability of the magnetic fields, but this can change in the future, as models improve over the years. Another advantage of our method is that it could allow in principle for the magnetic field model to be reverse-engineered from the constraints, instead of it being assumed. 

We use the Bayestar map \citep{Green2015, Green2018, Green2019}\footnote{the maps can be obtained using the Python dustmaps package, and can be visually inspected at \url{http://argonaut.skymaps.info/}}, the most recent iteration of which was created using parallaxes from the Gaia mission \citep{Collaboration2016b}. To create the map, stellar photometry from Pan-STARRS 1 \citep{Chambers2016} and 2MASS \citep{Skrutskie2006} was used.  We use this map over other available 3D reddening maps such as \citep{Leike2019} due to its distance depth of few kpc.


In this work we show the 3D temperature map can reconstruct the dust emission data satisfactorily, and that enough information is available to make temperature measurements along the line of sight to clouds like Cepheus. While we do not know if the model is accurate, there is enough constraining power for it to be precise, at least in areas with enough dust. The analysis approach also allows us to test whether the conversion factor between dust optical depth and reddening is a constant, as used in \cite{Schlegel1998}(hereafter, SFD) and the Planck dust map \cite{PlanckCollaboration2014}.
 
This paper is structured as follows: in Section \S \ref{sec:dust_3D_data} we describe the data used in the analysis, including Bayestar and the emission maps from \emph{Planck} and SFD/\emph{IRAS-DIRBE}. Section \S \ref{sec:dust_3D_methods} shows the mathematical model and the methods that we use in the analysis. Section \S \ref{sec:3D_dust_results} describes the results, and Section \S  \ref{sec:3D_dust_applications} discusses future applications of the 3D dust temperature map. Finally, we conclude in Section \S \ref{sec:dust_3D_conclussion}.
\section{Data}\label{sec:dust_3D_data}
\subsection{3D Dust Extinction Map}

Bayestar provides a 3D reddening map in 120 distance bins, evenly spaced in distance modulus from $\mu=4-19$ ($60$ pc to $60$ kpc) in steps of $\Delta\mu=0.125$ (Fig. \ref{fig:bayestar_distances}. In practice the maximum reliable distance is usually less than 8 kpc at high latitude, and $2-4$kpc at lower latitude where dust obscuration reduces the number of stars observed at greater distances. The uniform binning in $\mu$ is a natural choice, since the resolution of the map is closest to being uniform in log distance.

The pixel size of the original map varies across the sky, from HEALPix Nside 256 to 1024, depending on the local stellar density. In our analysis, we sample the entire sky at Nside 1024 (pixel size of $3.4'$) so in some areas the numbers are just a rebinning from Nside 256 or 512 to 1024.

Bayestar provides the reddening data stored as samples of cumulative reddening along each line of sight from the posterior modeling done by \cite{Green2019}. We retrieve all of the samples of cumulative reddening, take the median over samples, and then differentiate that map along each line of sight to obtain a reddening in each distance slice. We use the resulting ``median Bayestar'' map in the following analysis.

\subsection{Dust Emission Data}{\label{sec:planck_data}}
We synthesize dust emission maps over the entire sky with a resolution comparable to \emph{Planck}. We make use of the full-sky 217, 353, 545, 857GHz emission maps from the \emph{Planck} \citep{Tauber2010} satellite\footnote{The \emph{Planck} maps are obtained from  the PR4-2021 release at \url{https://pla.esac.esa.int/##mapshttps://pla.esac.esa.int/##maps}, selecting single frequency tab and downloading the full sky maps.} and \emph{IRAS/DIRBE} \citep{Schlegel1998} \footnote{The IRAS data could be obtained from \url{https://lambda.gsfc.nasa.gov/product/iras/}. However, IRAS got only 97$\%$  of the sky. \citep{Schlegel1998} backfilled the missing stripes with DIRBE data, as well as combined the IRAS and DIRBE maps to take advantage of the higher angular resolution of IRAS and the improved control on absolute calibration from DIRBE. Thus we utilized their map.} at 3000GHz. The 2048 Nside emission maps are rebinned to 1024 to match the extinction map. 

%
%

To obtain the dust emission maps,
we subtract from these maps the CMB anisotropy, and obtain the maps given in Fig. \ref{fig:dust_3D_emissison_data}. We are able to use this procedure to estimate the dust contribution to each frequency because at these frequencies ISM dust is the dominant foreground component \citep{Dwek1997, Collaboration2014b},
with the exception of the cosmic infrared background (the total dust emission from all unresolved galaxies along the line of sight). We do not have a good way of determining and removing the contribution of the CIB, so high-latitude regions of our map (with very little ISM dust) do not provide meaningful constraints on the temperature. However, methods like Generalized Needlet Internal Linear Combination \citep[GNILC;][]{Remazeilles2011} might be useful to disentangle the different foregrounds. In this work we focus on intermediate latitudes where the ISM dust is sufficiently bright that it is reasonable to assume it dominates the CIB.









\section{Methods}\label{sec:dust_3D_methods}

In this section we describe the method used to reconstruct the emission map from the 3D extinction map. We present our model in Section \S \ref{sec:dust_3D_model}, the smoothing and masking of the maps in Sections \S \ref{sec:dust_3D_smoothing} and \ref{sec:dust_3D_mask}, and the fitting and sampling methods in Sections \S \ref{sec:dust_3D_optimizer} and \ref{sec:dust_3D_sampling}. 

\begin{figure*}[t!]
 \hspace{-0.5cm}
	\includegraphics[width=1.05\textwidth]{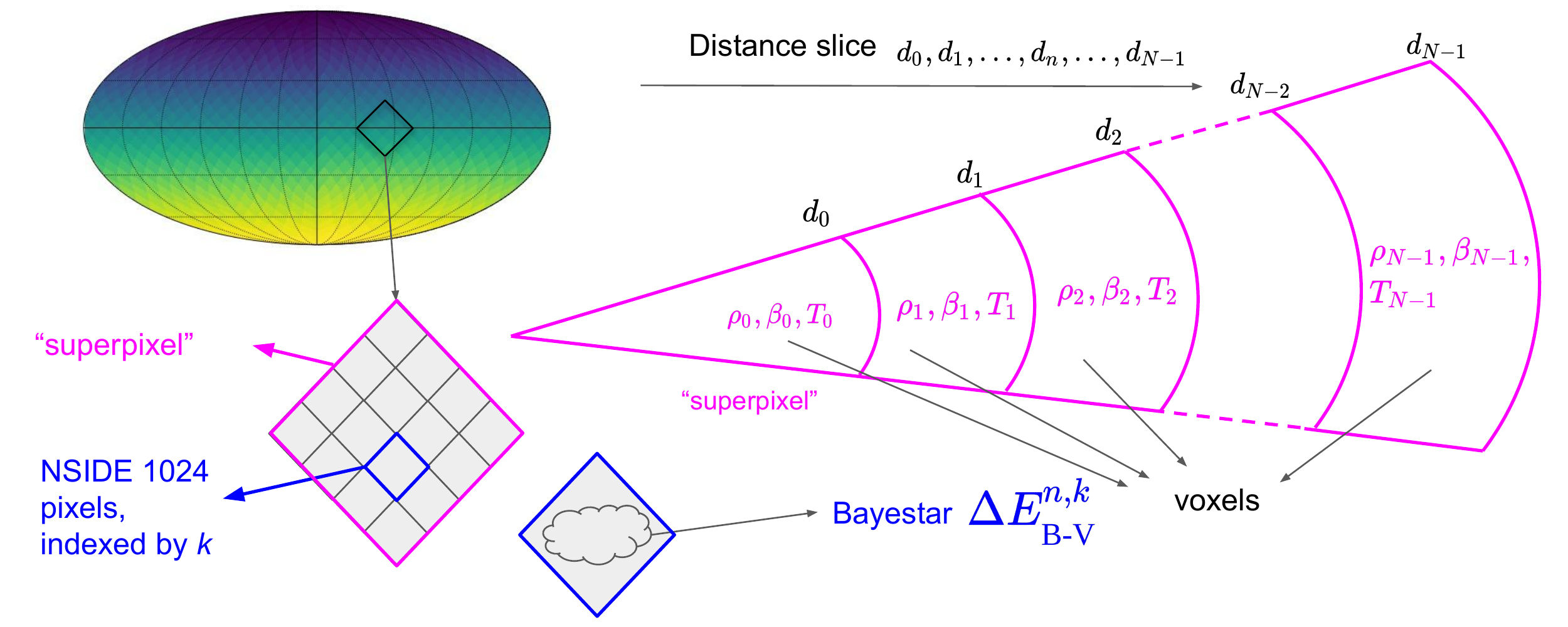}
	\caption[Description of the Superpixel Fit Method]{Description of the superpixel fit method. The maps containing the extinction and emission data are at a resolution of Nside 1024. The dust parameters map at a lower resolution, necessary to constrain the fit; a pixel in the parameter map, called a ``superpixel'', is overlayed over multiple pixels from the data. The voxel parameters take a single value for the entire area of the superpixel. \label{fig:superpixel_method}}
\end{figure*}
\subsection{Model}\label{sec:dust_3D_model}
Our base model relies on using the Nside 1024 HEALPix reddening map and emission maps to create maps of dust emission parameters ($T, \beta, \rho$) of reduced Nside, such as Nside 32, 64, or 128, but letting these dust parameters vary along the line of sight.  We call a HEALPix pixel in parameter maps a ``superpixel'', because the parameters of the superpixel are determined by the information in the subpixels corresponding to it in the higher resolution data maps.  Because each superpixel contains many Nside 1024 pixels (e.g., an Nside 64 superpixel contains 256 of them), the dust parameters are constrained as long as there is enough dust structure within the superpixel.

For each angular superpixel, we consider a series of $N$ total distance slices $d_0, d_1, ..., d_{N-1}$ (see Fig. \ref{fig:superpixel_method}). The distance slice series is indexed by $n$, ${d_n}$. They delimit ``voxels'', with voxel $n$ bounded by inner and outer distance slices $d_{n-1}$ and $d_n$ along the line of sight, with the exception of the first voxel, which extends from the observer to the first distance slice $d_0$.

The voxels each contain a given amount of dust, whose emission, when summed over all the voxels along a line of sight, should reproduce the dust emission data in 2D from \emph{Planck} and \emph{IRAS}.

In each voxel, we model dust emission as a single modified black body (MBB):
\begin{equation}\label{eq:MBB_ISM_dust}
\Delta I_{\nu}^{\textrm{voxel}~n,~k} = \tau_{353}^{n,k} \big ( \frac{\nu}{\nu_0} \big )^{\beta^n} B_{\nu}(T^n) 
\end{equation}
where $n$ is the index of the voxel slice (with possible values from $0$ to $N-1$), $\nu_0=353$GHz, $\beta^n$ and $T^n$ are the MBB power law index and the dust temperature in voxel $n$, $B_{\nu}(T^n)$ is the Planck blackbody spectral radiance for $T^n$, and $\tau_{353}^n$ is the optical depth normalized to 353GHz. Here we make the approximation that dust emission is optically thin, $1-\exp^{-\tau} = \tau$ is a good approximation, and therefore we can do the sum along the line of sight.

The dust optical depth $\tau_{353}$ can be connected to the dust reddening $E_{\textrm{B-V}}$ using a conversion factor $\rho_{353}$, in the following way:
\begin{equation}\label{optical_depth}
\tau_{353}^{n,k} = \rho_{353}^n \Delta E_{\textrm{B-V}}^{n,k}
\end{equation}
where $k$ is indexing over the pixels from the data maps contained in the given superpixel, and $n$ is indexing over the voxels along the line of sight.

We do the fit in $\sim 2^{\circ} \times 2^{\circ}$  superpixels (Nside=32) to  \emph{Planck} 217, 353, 545, 857 GHz and SFD $3000$ GHz. We assume that for the sightlines contained within a superpixel, within each voxel $n$, $\rho_{353}^n,\beta^n,T^n$ are constant, and thus don't have a dependence on $k$.


Summing over the voxels delimited by each distance slice and superpixel angular boundaries, we obtain:

\begin{equation}
I_{\nu}^{\textrm{total},k}= O_{\nu} + \sum_n  \Delta I_{\nu}^{\textrm{voxel}~n,~k}
\end{equation}

\begin{equation}
I_{\nu}^{\textrm{total},k}= O_{\nu} + \sum_n \rho_{353}^n \Delta E_{\textrm{B-V}}^{n,k} \big ( \frac{\nu}{\nu_0} \big )^{\beta^n} B_{\nu}(T^n), 
\end{equation}

where $O_{\nu}$ are (full-sky) offsets that can be included or omitted in the fit, to account for potential offsets in the data. Here the index $\nu$ is used to refer to the frequency indexing the emission data sets used in the analysis.

\subsubsection{Bayesian Inference}\label{sec:dust_3D_bayesian inference}

Bayes' Theorem relates the data D with the set of parameters of the model $\{\theta\}$:

\begin{equation}\label{eq:bayes}
p(\theta|\textrm{D})p(\textrm{D})= p(\textrm{D}|\theta)p(\theta)    
\end{equation}
The posterior probability of the parameters given the data is:
\begin{equation}\label{eq:posterior}
p(\theta|\textrm{D}) = \frac{p(\textrm{D}|\theta)p(\theta)}{p(\textrm{D})}
\end{equation}
In our case, the data are
$\{I_{\nu}^{\textrm{D},k}\}$,$\{\sigma_{\nu}^{\textrm{D}}\}$,$\{\Delta E_{\textrm{B-V}}^{k,n}\}$, and the parameters are $\{\rho_{353}^n\},\{\beta^n\},\{T^n\}, \{O_{\nu}\}$.

The likelihood of the data given the parameters is:
\begin{equation}\label{eq:ch3_likelihood}
\begin{split}
\mathcal{L} &=p(\textrm{D}|\theta)\\
            &= p(\{I_{\nu}^{\textrm{D},k}\},\{\sigma_{\nu}^{\textrm{D}}\},\{\Delta E_{\textrm{B-V}}^{k,n}\}|\{\rho_{353}^n\},\{\beta^n\},\{T^n\})\\
            &=\prod_{k}\prod_{\nu} \frac{1}{\sqrt{2\pi{\sigma_{\nu}^{\textrm{D}}}^2}} \exp{-(I_{\nu}^{\textrm{total},k}-I_{\nu}^{\textrm{D},k})^2/2{\sigma_{\nu}^{\textrm{D}}}^2}
\end{split}
\end{equation}

We can write the log likelihood as  $\ln{\mathcal{L}} =K -0.5 \times \Delta \chi^2$, with $K$ being a constant not depending on the parameters, but only on the data.
\begin{equation}{\label{eq:chi_square}}
\Delta \chi^2 = \sum_{k} \sum_{\nu} \left ((I_{\nu}^{\textrm{total},k}-I_{\nu}^{D,k})/\sigma_{\nu}^{\textrm{D}}\right)^2,
\end{equation}
where $k$ is indexing over the sightlines in the superpixel, $I_{\nu}^{\textrm{reference}}$ is the dust intensity coming from subtracting the CMB from \emph{Planck} data, as described above in Section \S \ref{sec:planck_data}, and $\Delta I_{\nu}^{\textrm{total}}$ is the intensity obtained along the line of sight for each pixel in the superpixel.

\begin{figure*}[th!]
	\begin{center}
	\begin{tabular}{cccc}
    	\textbf{(a)}  & \textbf{(b)} \\[6pt]
		\hspace{-0.20mm}
		\includegraphics[width=0.48\textwidth]{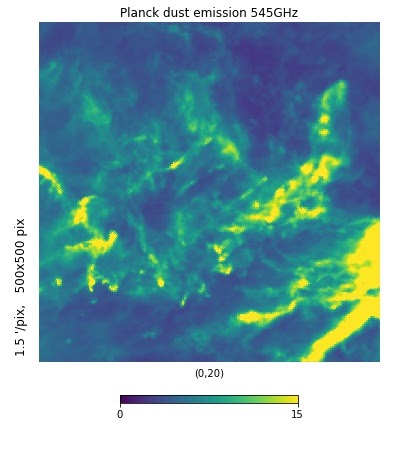} &
		\hspace{-1.70mm}	
		\includegraphics[width=0.48\textwidth]{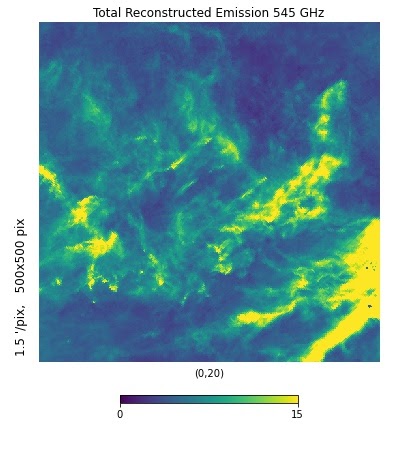} \\
	\end{tabular}
	\vspace{-4.00mm}

	\begin{tabular}{cccc}
	    \textbf{(c)}  & \textbf{(d)} \\[6pt]
		\hspace{-0.20mm}
		\includegraphics[width=0.48\textwidth]{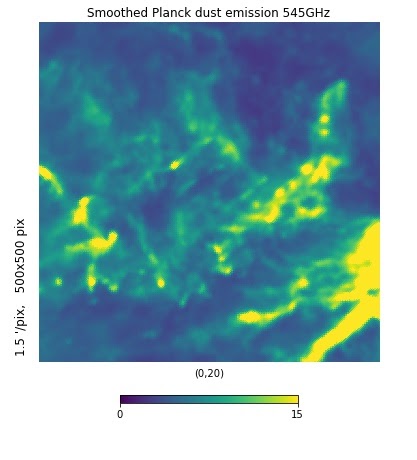} &
		\hspace{-1.70mm}		
		\includegraphics[width=0.48\textwidth]{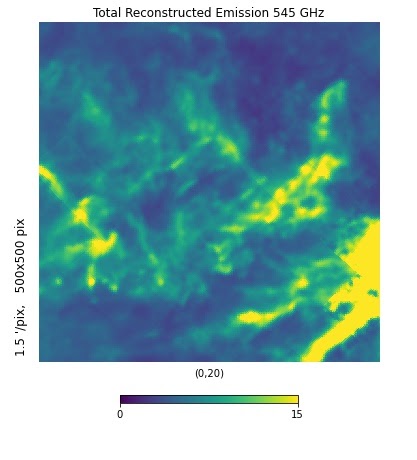} \\
		\vspace{-5.00mm}
		\vspace{-2.00mm}
		
	\end{tabular}
	\end{center}

	\caption[Smoothed Maps]{ (a) The data for dust emission at 545 GHz, after the processing described in Section \S \ref{sec:dust_3D_data}. The plot is zoomed in on the sky area of the $\rho$ Ophiuchi cloud complex. (b) The reconstruction of the emission data in (a) using the 3D dust reddening fits. We noticed that even though the images match well in general, the reconstruction has higher resolution than the emission data. This is due to Bayestar having higher resolution than the \emph{Planck} maps. Thus, smoothing is required to match the "point spread functions" of the emission and reddening data. (c) Data emission map after smoothing to 10'. (d) Reconstructed emission map based on both emission data and Bayestar smoothed to 10'. Figure \ref{fig:difference_smoothed_maps} shows the difference between maps (c) and (d).}\label{fig:smoothed_maps}
\end{figure*}

\begin{figure}[th!]
	\begin{center}
    \includegraphics[width=0.45\textwidth]{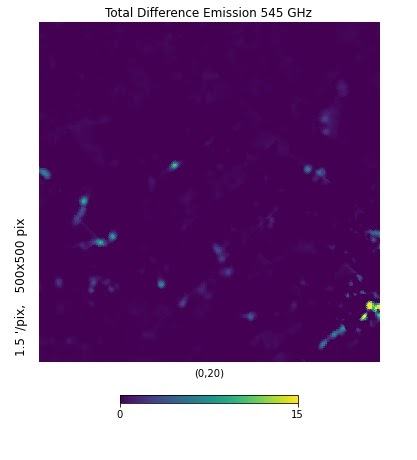} 
	\caption[Difference Between Smoothed Data and Reconstructed Maps]{The difference between the smoothed emission data and the reconstruction based on the smoothed reddening map ( Figs (c) and (d) in \ref{fig:smoothed_maps}. If the relative point spread functions would not match, we would see contours appear around brighter spots; the fact that we don't see them is an indication the smoothing was successful.}
	
	\label{fig:difference_smoothed_maps}
	\end{center}
\end{figure} 

\paragraph{Priors}
We impose priors on the parameters that are allowed to float along the line of sight. This is necessary due to the voxels that contain little to no dust, and in which the sampler would otherwise not be able to constrain the parameters. For the case when the temperature is the only parameter varying along the line of sight, the prior $p(\theta)$ is given by

\begin{equation}\label{eq:prior}
p(\{\rho_{353}^n\},\{\beta^n\},\{T^n\}) = \prod_{n} \frac{1}{\sqrt{2\pi{\sigma_{T}^2}}} \exp{-\frac{(T^n-T_0)^2}{2{\sigma_{T}}^2}}
\end{equation}
with $T_0 =18.1$K, $\sigma_{T}=9$K 

\begin{equation}
    \ln{p(\{\rho_{353}^n\},\{\beta^n\},\{T^n\})}=K_1 -\frac{1}{2}\sum_{n}\frac{(T^n-T_0)^2}{{\sigma_{T}}^2}
\end{equation}

\begin{figure*}[th!]
	\centering
	\begin{tabular}{cccc}
    	\textbf{(a)}  & \textbf{(b)} \\[6pt]
		\hspace{-2.90mm}
		\includegraphics[width=0.49\textwidth]{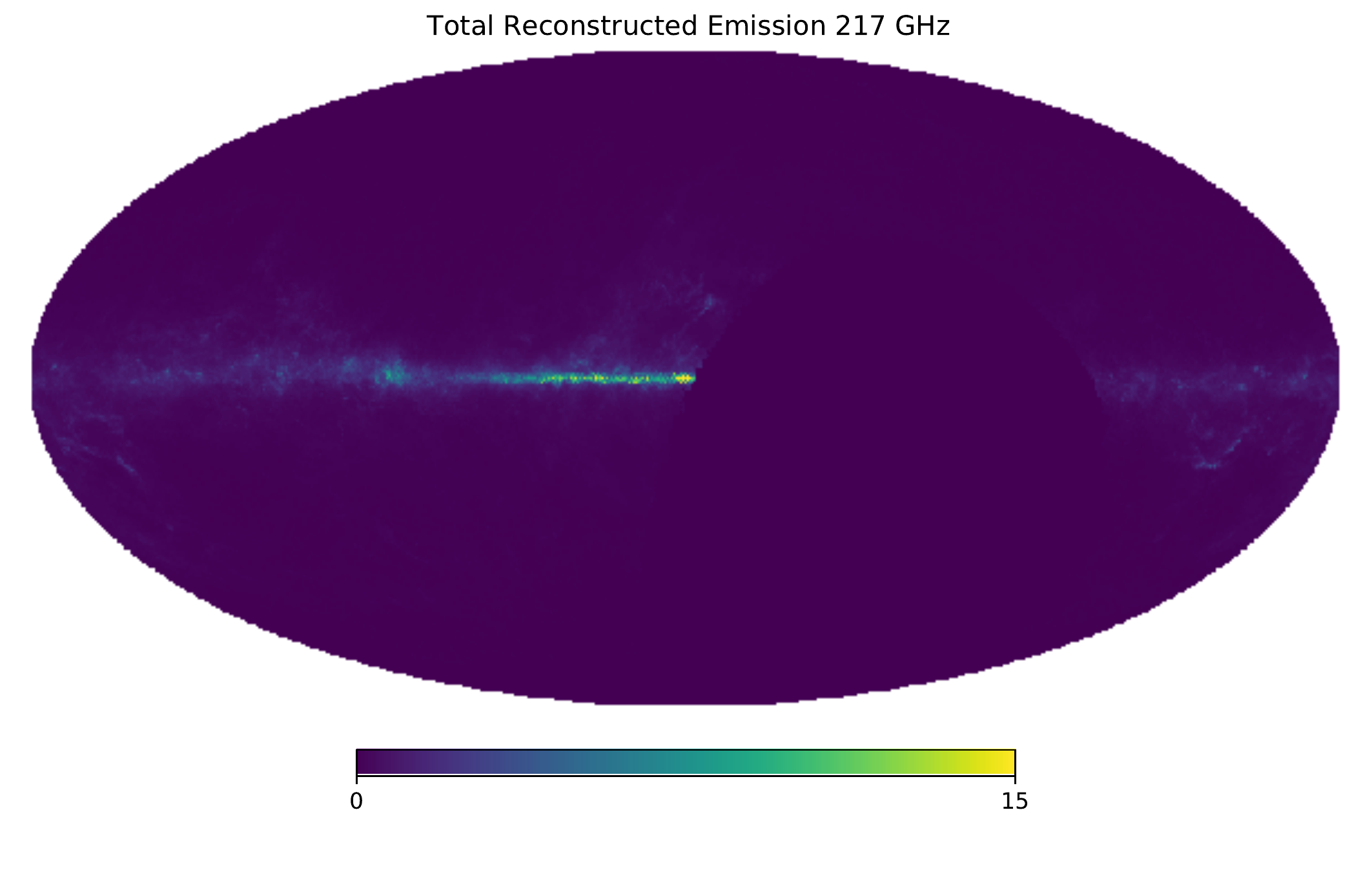} &
		\hspace{-3.00mm}	
		\includegraphics[width=0.49\textwidth]{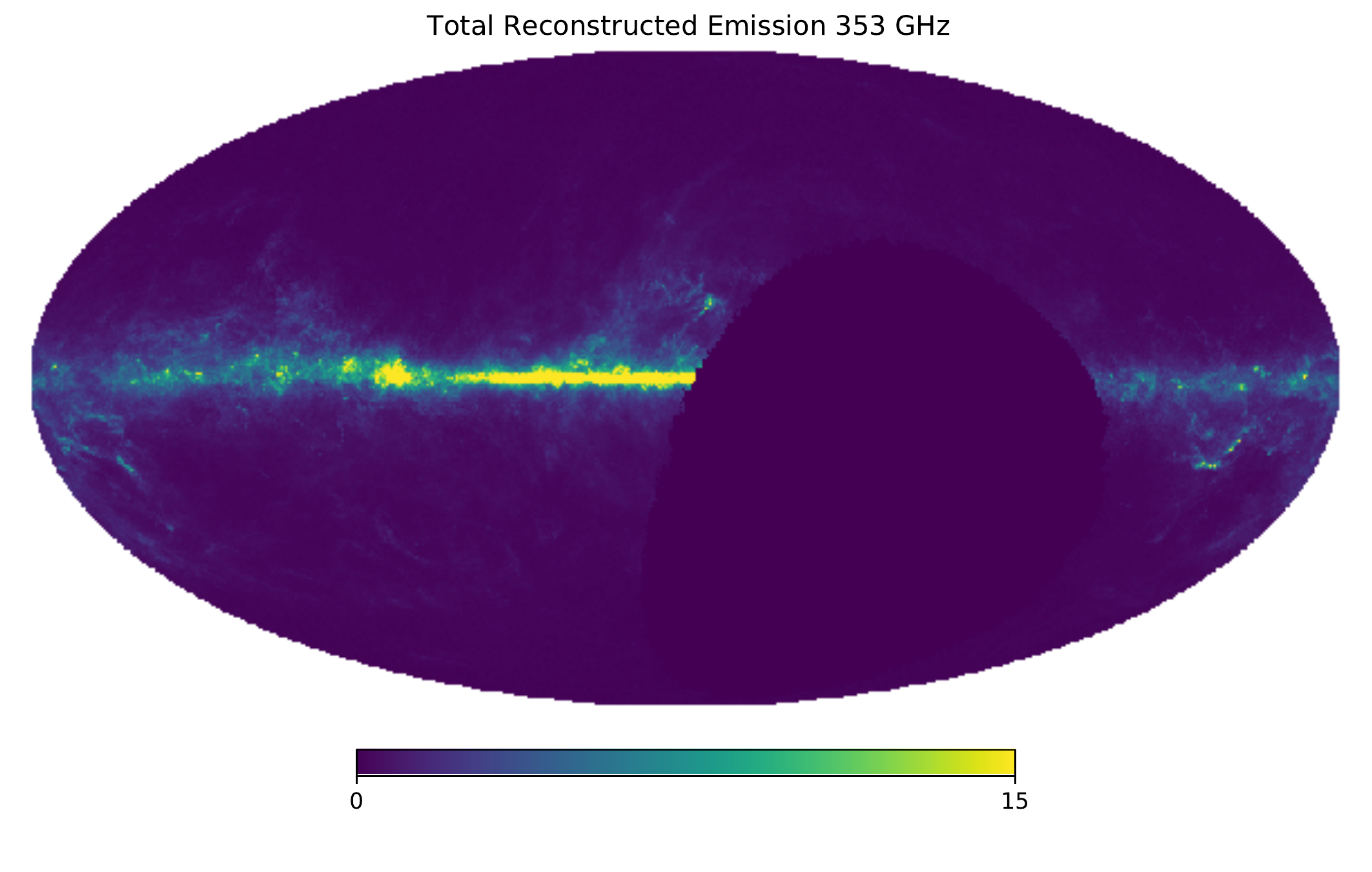} \\
	\end{tabular}
	\vspace{-4.00mm}
    \centering

	\begin{tabular}{cccc}
    	\textbf{(c)}  & \textbf{(d)} \\[6pt]
    	\hspace{-2.90mm}
    	\includegraphics[width=0.49\textwidth]{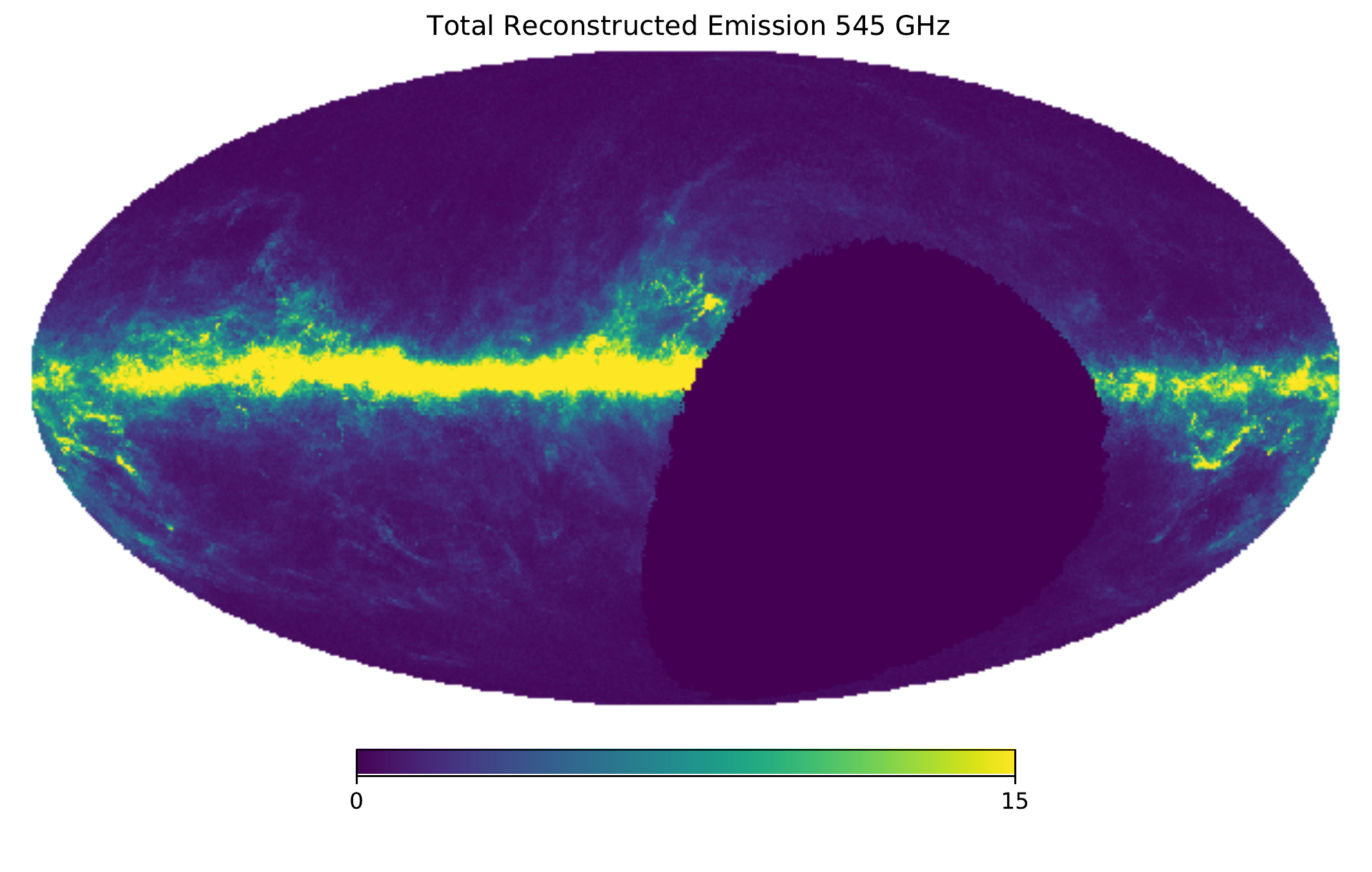} &
    	\hspace{-3.00mm}	
    	\includegraphics[width=0.49\textwidth]{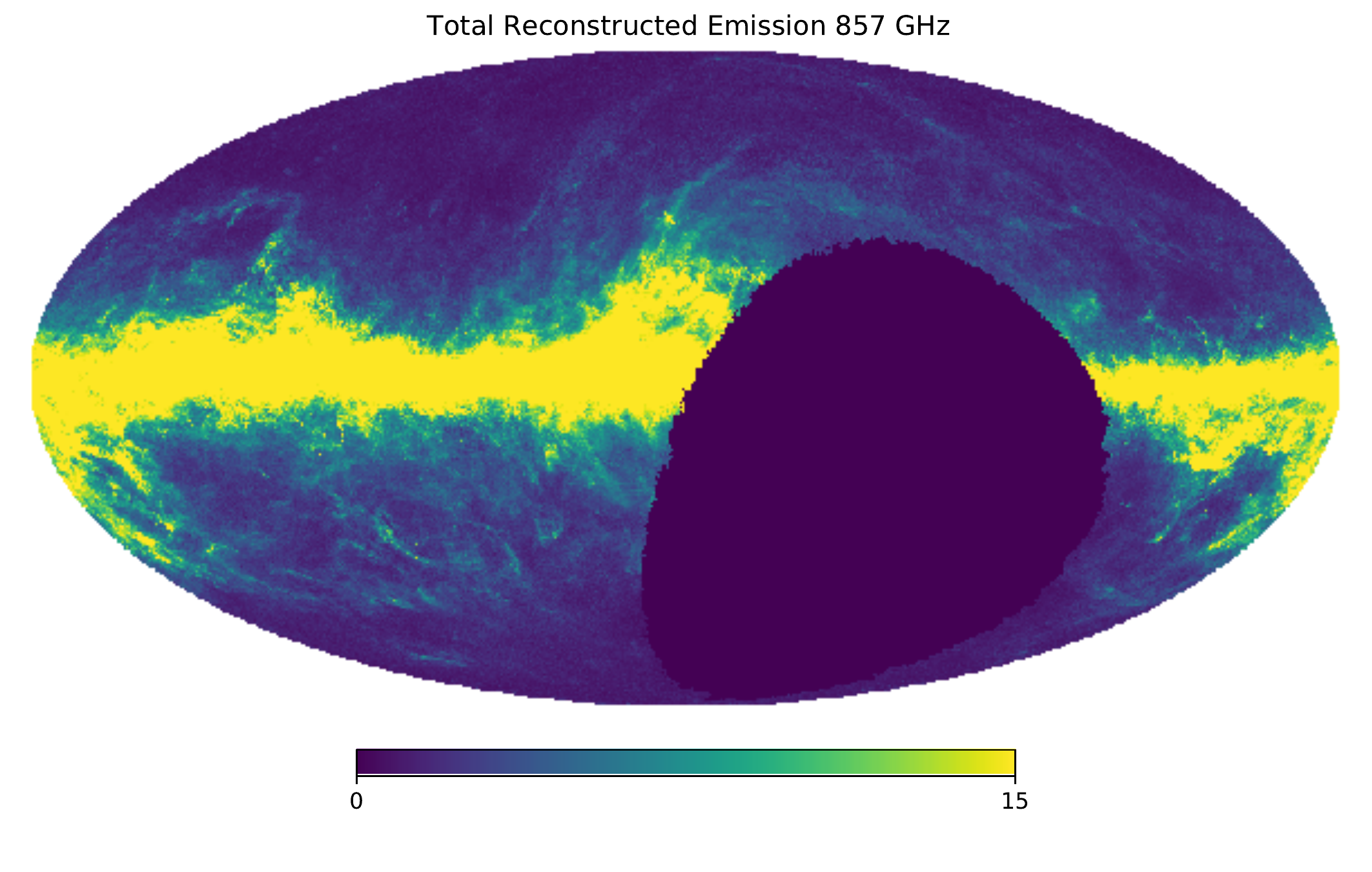} \\
    \end{tabular}
    \vspace{-4.00mm}
    \centering

	\begin{tabular}{cccc}
    	\textbf{(e)}  & \textbf{(f)} \\[6pt]
    	\hspace{-2.90mm}
    	\includegraphics[width=0.49\textwidth]{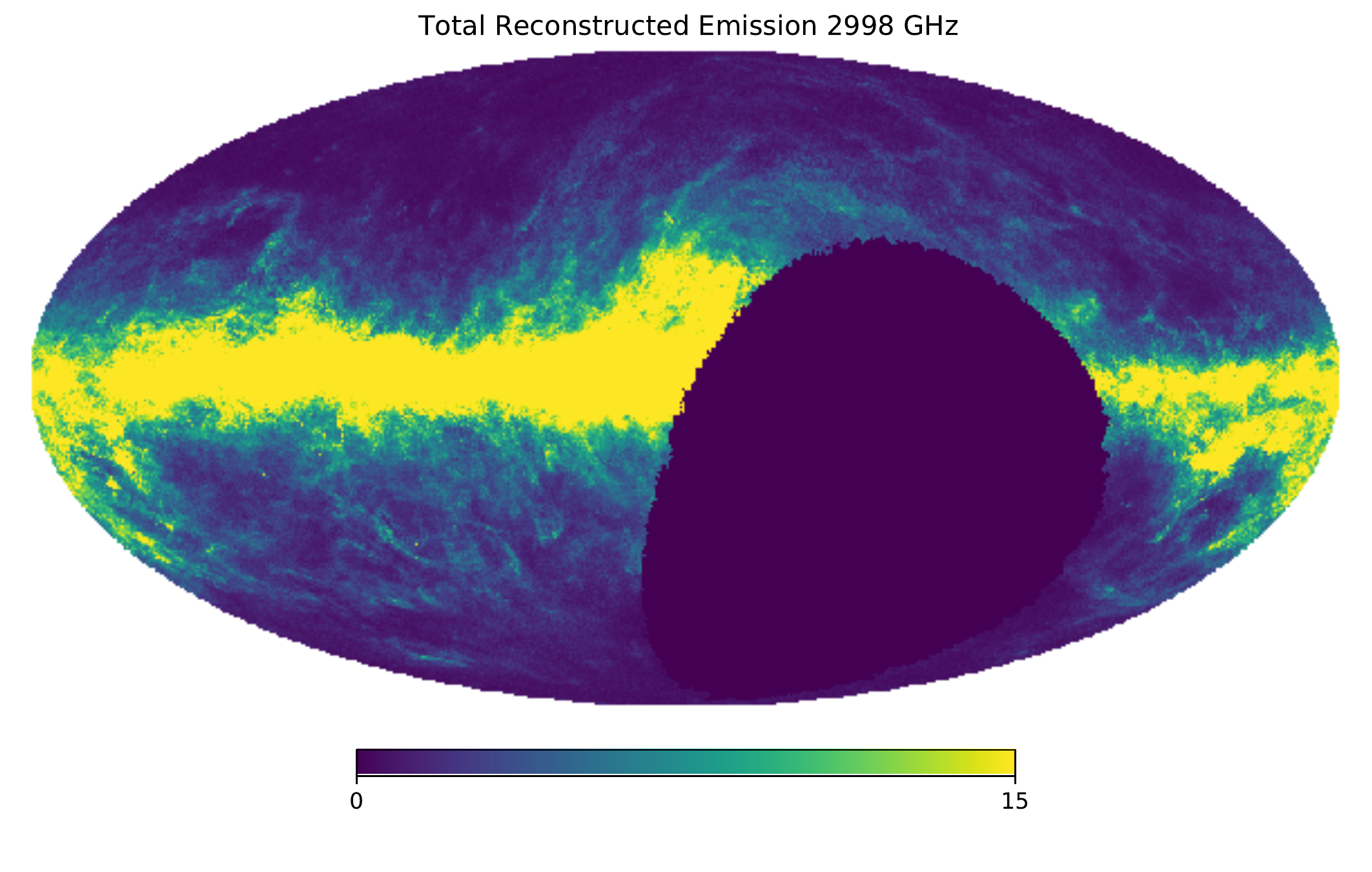} &
    	\hspace{-3.00mm}	
    	\includegraphics[width=0.49\textwidth]{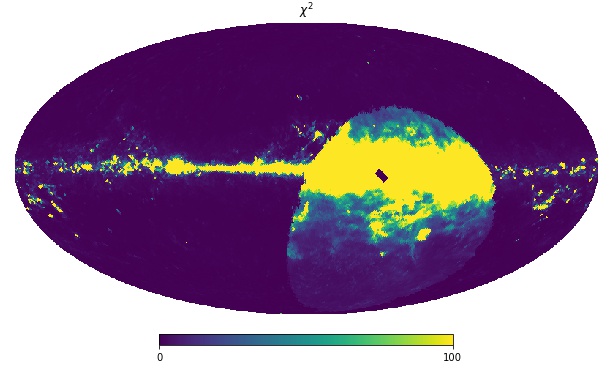} \\
    \end{tabular}

	\caption[Reconstructed Dust Emission Sky Maps]{ (a), (b),(c),(d),(e): Reconstructed dust emission sky maps. Comparing with Fig. \ref{fig:dust_3D_emissison_data}, we see that it is possible to reconstruct the dust emission maps from the 3D reddening maps. (f) The $\chi^2$ for each superpixel fit. While the scale of the $\chi^2$ depends on the error analysis, this results indicate a good fit for most of the sky, with the exception of the regions closer or in the galactic plane with a lot of dust, where Bayestar may not be modeling the dust beyond a certain distance. }
	
	\label{fig:reconstructed_dust_emission_sky_maps}
\end{figure*}

\subsection{Smoothing}\label{sec:dust_3D_smoothing}
The emission and reddening maps have different resolutions, and it is important when performing the fit to have the maps match (see Fig. \ref{fig:smoothed_maps}). 
The 217 GHz \emph{Planck} map has a point spread function (PSF) of 5.5', the 353, 545, 857 GHz \emph{Planck} maps have a PSF of about 5' each, and the SFD/\emph{IRAS-DIRBE} map has a PSF of 6.1'.
Bayestar does not have a well defined PSF because each angular pixel depends on a disjoint set of stars.  There is some regularization in the map, but on a physical distance scale, not in the form of an angular PSF \citep{Green2019}. We take the Bayestar FWHM to be the pixel size of the HEALPix Nside 1024 map, which is 3.4'. We smooth all maps to a PSF FWHM of 10'. 

Thus, we smooth the 217 GHz \emph{Planck} map with a Gaussian convolution \citep{Gorski2005, Zonca2019} of 8.35', the 353, 545, 857 GHz \emph{Planck} maps with one of 8.66', the \emph{SFD/IRAS-DIRBE} with 7.92', and Bayestar with 9.39'. The resulting maps appear to match well (Figs. \ref{fig:smoothed_maps}  and \ref{fig:difference_smoothed_maps}).


\begin{figure*}[t]
\hspace{-4mm}
	\includegraphics[width=1.05\textwidth]{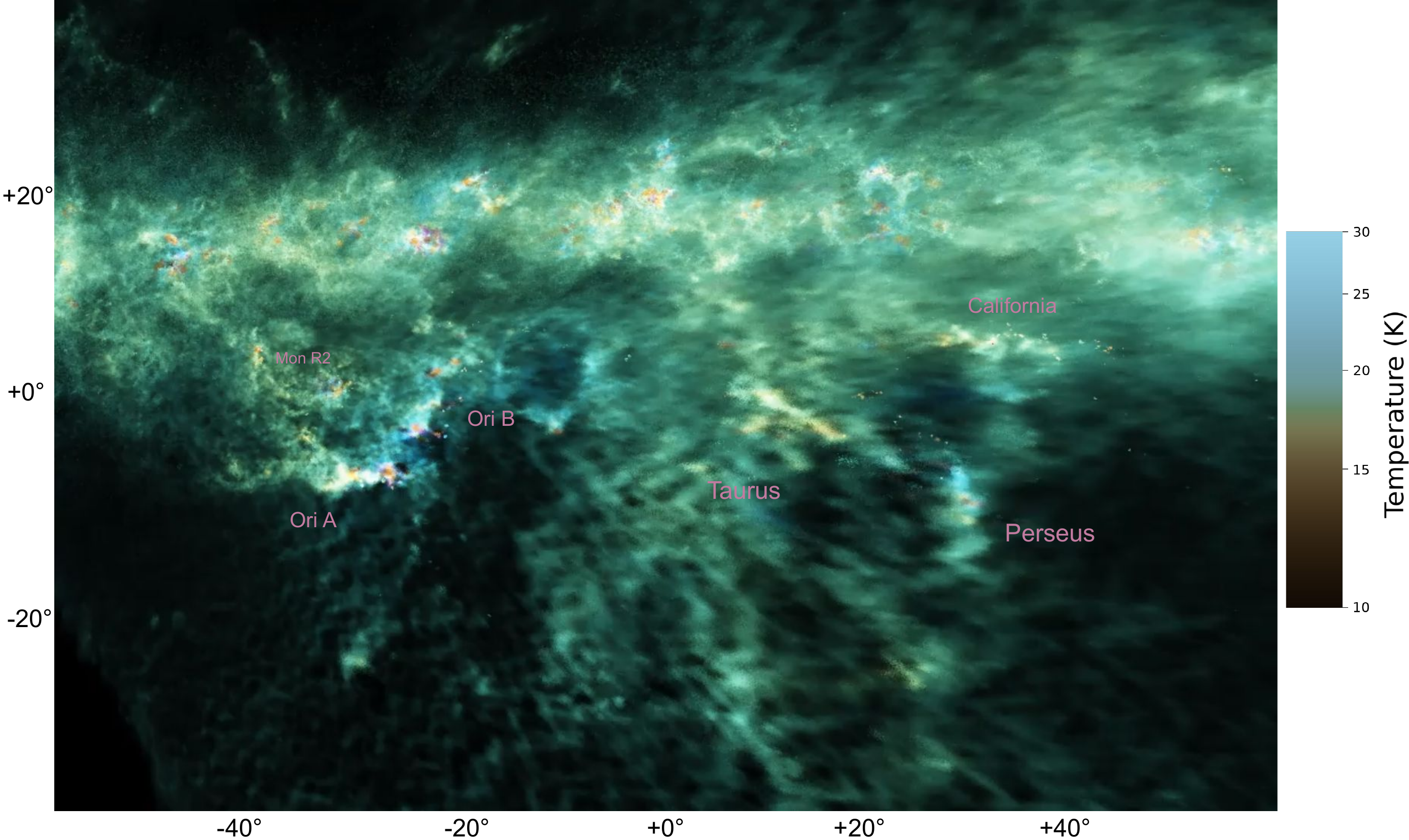}
	\caption[3D Map Visualization]{We generate 3D visualizations 
	of the Nside 128 ($27'$ resolution) map of the temperature of galactic dust and its density. The intensity is proportional to the density, and the RGB variation is temperature dependent. Blue is sensitive to the shortest wavelength (hotter dust, ~20K), red to longer wavelengths (cold dust), and green to intermediate values. The perspective shown here features a screenshot the camera placement doing a 25-pc loop around the Sun, looking in the galactic plane towards the anti-center (180$^{\circ}$ galactic longitude), towards the Orion, Taurus, Perseus, and California clouds. 
	The videos can be seen at the authors' \href{https://ioanazelko.com/3d-galactic-dust-temperature-map/}{personal websites}, as well as on \href{https://youtube.com/playlist?list=PLijDPBNhGIqQccgN0BKJciE16NJ1PtsdO}{Youtube}. 	\label{fig:3D_map}}
\end{figure*}

\subsection{Masking}\label{sec:dust_3D_mask}
There are a few criteria used to select pixels from the extinction dataset to use in our analysis.
Bayestar provides a flag marking the convergence status of the pixels, and we include only converged pixels. Additionally, there is information provided regarding the minimum reliable distance. This is determined by finding the minimum distance slice that has at least three stars in front of it. We eliminate the pixels in which the minimum distance is infinity due to lack of any stars.

Another important aspect is the amount of dust present in the fit. At high $|b|$, there is very little reddening, which makes the three dimensional analysis impossible to constrain. Thus, the bin cuts need to be adjusted to have a single bin in those areas.
\subsection{Optimizer}\label{sec:dust_3D_optimizer}
We use an optimizer\footnote{from the Python package Scipy - Optimize} to fit the model described in Section \S \ref{sec:dust_3D_bayesian inference}. Due to its speed, we are able to fit the entire sky at HEALPix resolutions of up to Nside 128. For our problem, it is possible to compute the gradient and the Hessian matrix analytically, for each direction and distance cut choice in the sky. See Appendix \ref{sec:appendix} for the gradient and Hessian matrix calculations.

\subsection{Sampling}\label{sec:dust_3D_sampling}
While the optimizer allows us to find the maximum of the posterior, it does not give us information about the errors and actual posterior shape of the parameters. To get a sense of how precise our results are, we also use a sampler in a limited part of the sky as a check.

The sampler uses the \texttt{ptemcee} \footnote{The code can be found at the Python repository at \url{https://pypi.org/project/ptemcee/} or at the Will Vousden Github repository at \url{https://github.com/willvousden/ptemcee}} \cite{Vousden2016} package, which uses parallel tempering. This allows for a more efficient exploration of the parameter space than something like the Metropolis-Hastings algorithm.  We used 3 temperatures and 50 walkers, running for 30000 steps. The runs were thinned by a factor of 10 to reduce autocorrelation. To discard the burn-in phase of the chain, only the last 50\% of the steps were retained. To initialize the sampler, we first run the optimizer and use its results to generate the initial positions for the walker; this allows the sampler to converge faster, but also poses the risk of missing posterior modes if those modes are located far apart. However, for our model, it is not likely that the posteriors are multi-modal, based on tests we ran without an optimized initial starting point.

This method was applied to determine the temperatures of the Cepheus cloud, as described in Section \S \ref{sec:cepheus_cloud_temperature}.
\section{Results and Discussion}\label{sec:3D_dust_results}
\subsection{Reconstruction of the dust emission map from the 3D reddening map}

Our analysis's main intent is the show the proof-of-concept of the method of fitting a 3D dust reddening map to the 2D dust emission spectrum, which given future improvements in reddening maps, has the potential to achieve many goals.

\begin{figure*}[th!]
	\centering
	\begin{tabular}{cccc}
    	\textbf{(a)}  & \textbf{(b)} \\[6pt]
		\hspace{-3.20mm}
		\includegraphics[width=0.50\textwidth]{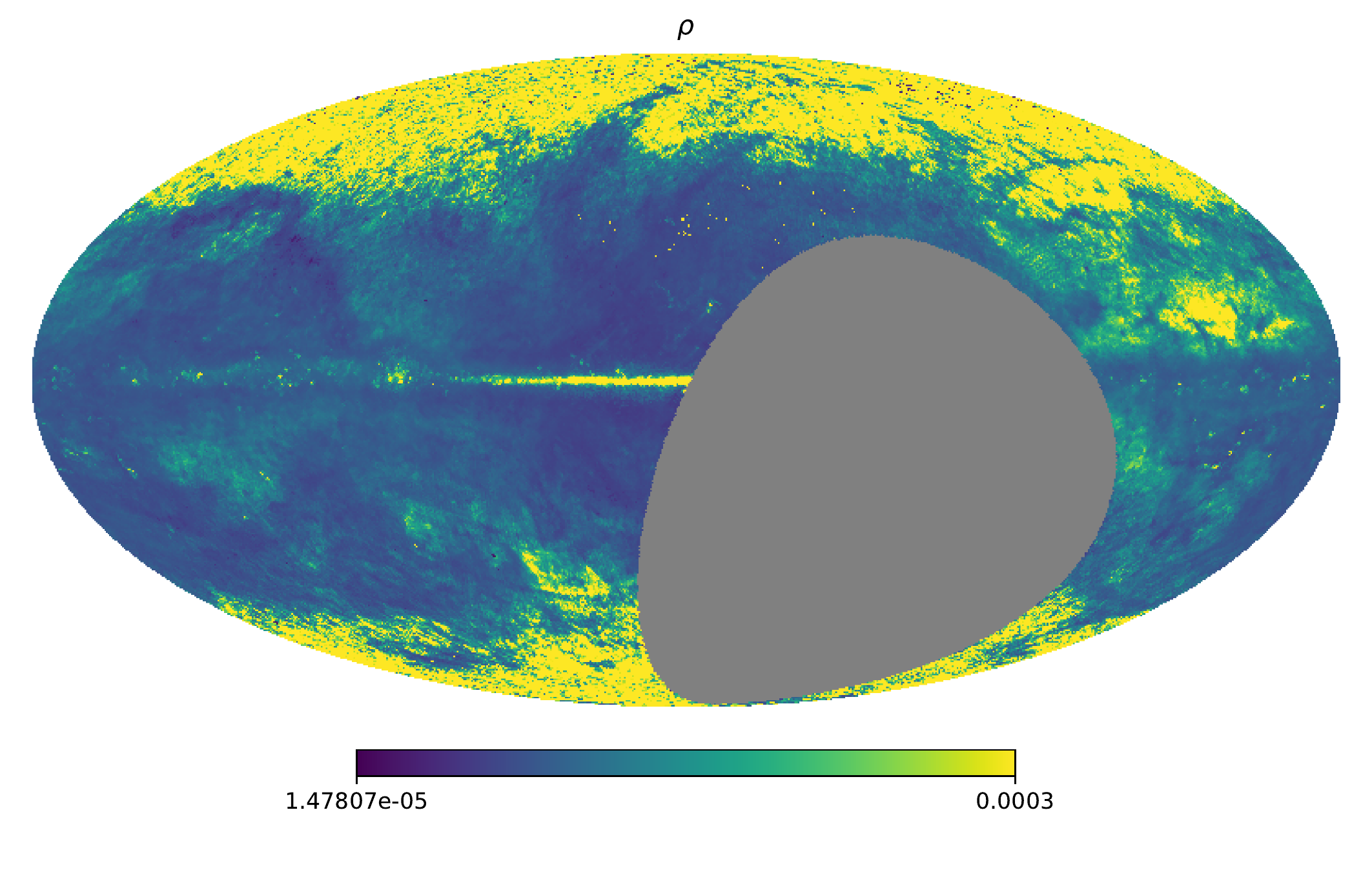} &
		\hspace{-2.70mm}	
		\includegraphics[width=0.50\textwidth]{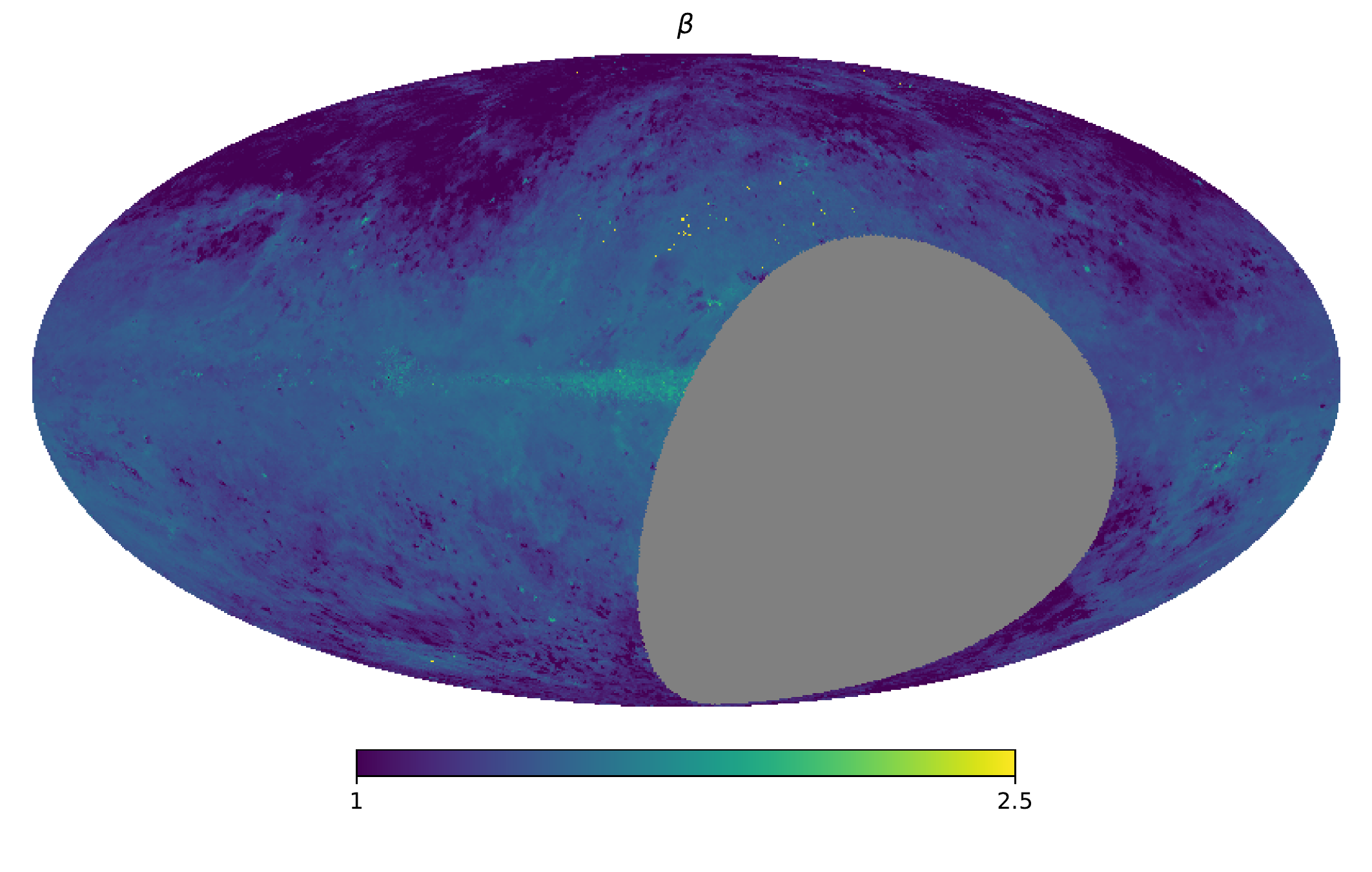} \\
	\end{tabular}
	\vspace{-4.00mm}

	\caption[Maps of the Conversion Coefficient $\rho$ and the Power-Law Index $\beta$]{ Maps of the conversion coefficient $\rho$ (left) and the power-law index $\beta$ (right). $\rho$ is shown to vary across the sky, and its variation would be useful to take into account for projects that rely on transforming emission maps into extinction maps. The large grey area corresponds to  missing data in the extinction map. }
	\label{fig:rho_and_beta}
\end{figure*}

\begin{figure*}[th!]
	\centering
	\begin{tabular}{cc}
		\hspace{-2.20mm}
		\includegraphics[width=0.50\textwidth]{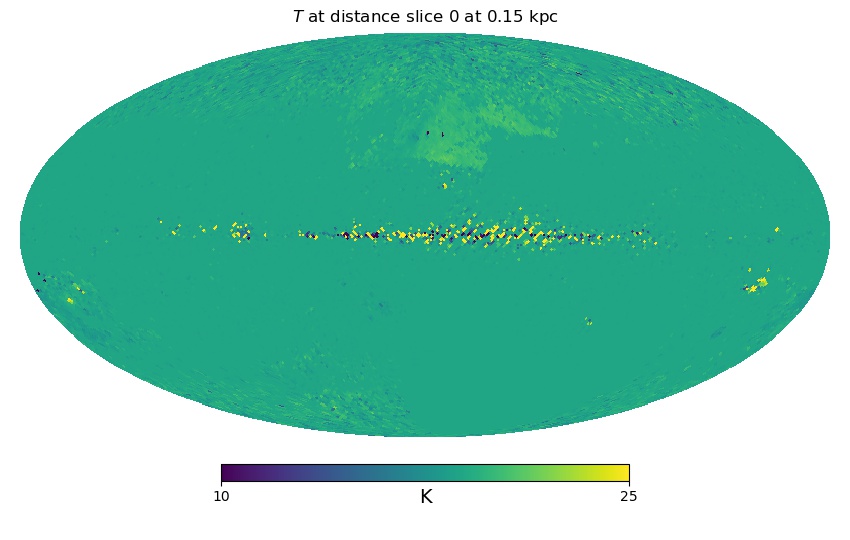} &
		\hspace{-4.70mm}
		\includegraphics[width=0.50\textwidth]{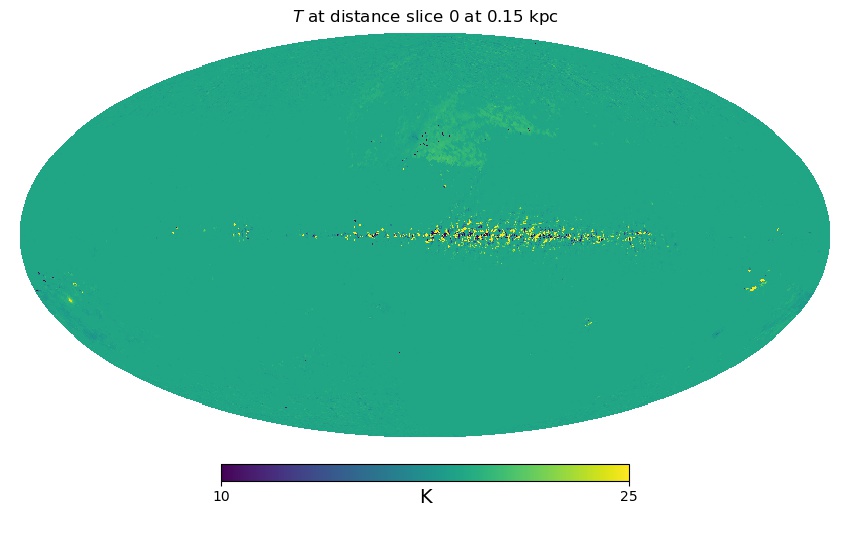} \\
	\end{tabular}
	
	\vspace{-10.00mm}
	
		\begin{tabular}{cc}
		\hspace{-2.20mm}
		\includegraphics[width=0.50\textwidth]{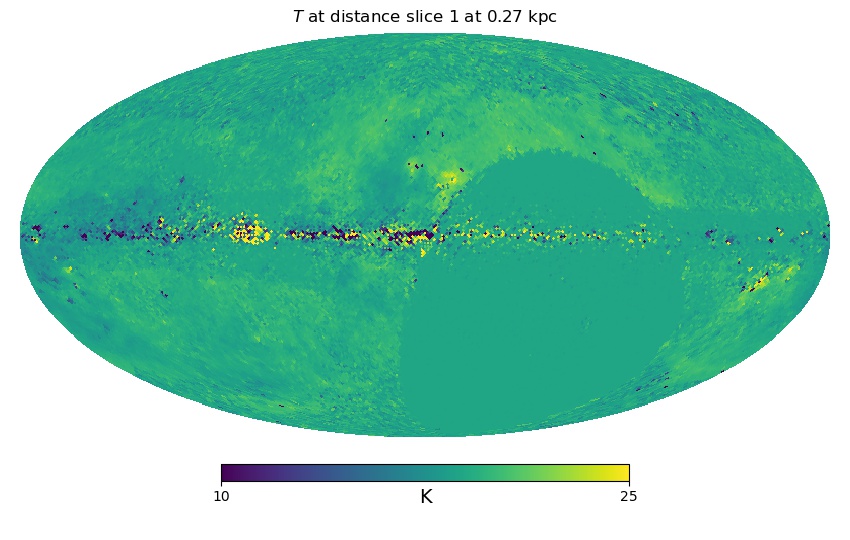} &
		\hspace{-4.70mm}
		\includegraphics[width=0.50\textwidth]{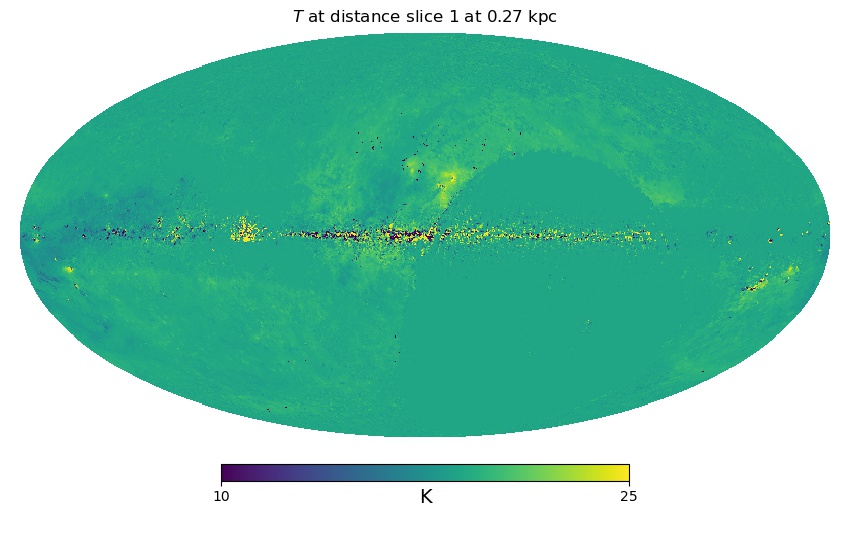} \\
	\end{tabular}
	
	\vspace{-10.00mm}
	
	\begin{tabular}{cc}
		\hspace{-2.20mm}
		\includegraphics[width=0.50\textwidth]{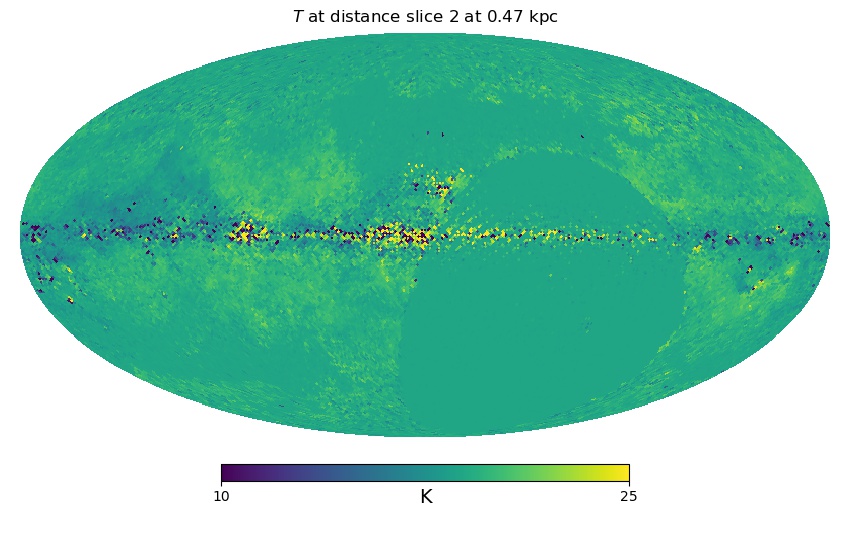} &
		\hspace{-4.70mm}
		\includegraphics[width=0.50\textwidth]{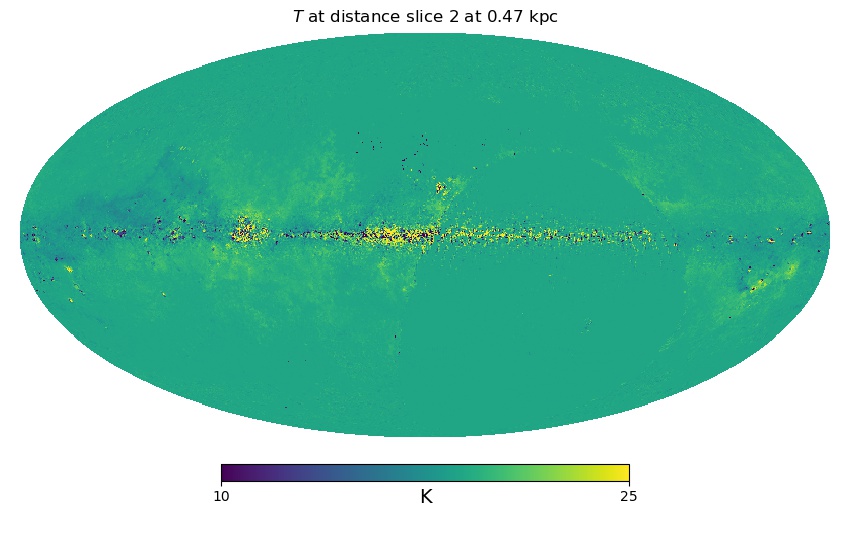} \\
	\end{tabular}
	
	\vspace{-10.00mm}
	
	\begin{tabular}{cc}
		\hspace{-2.20mm}
		\includegraphics[width=0.50\textwidth]{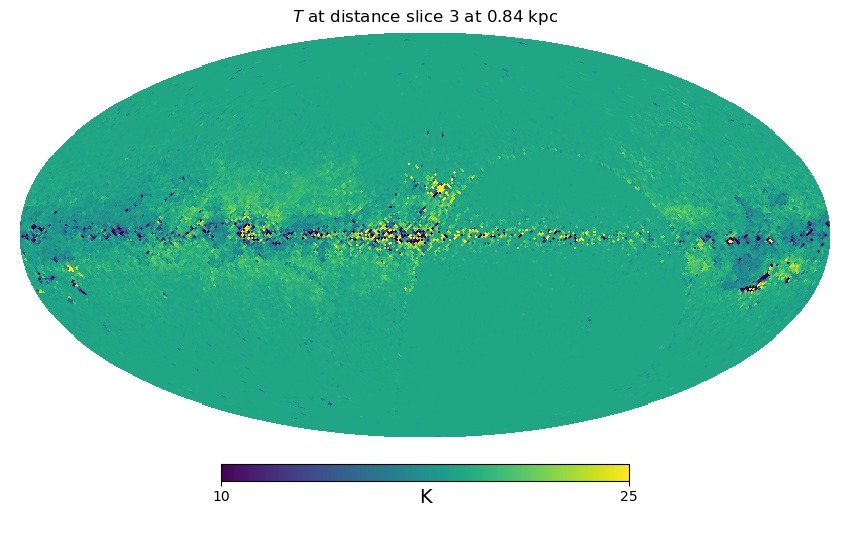} &
		\hspace{-4.70mm}
		\includegraphics[width=0.50\textwidth]{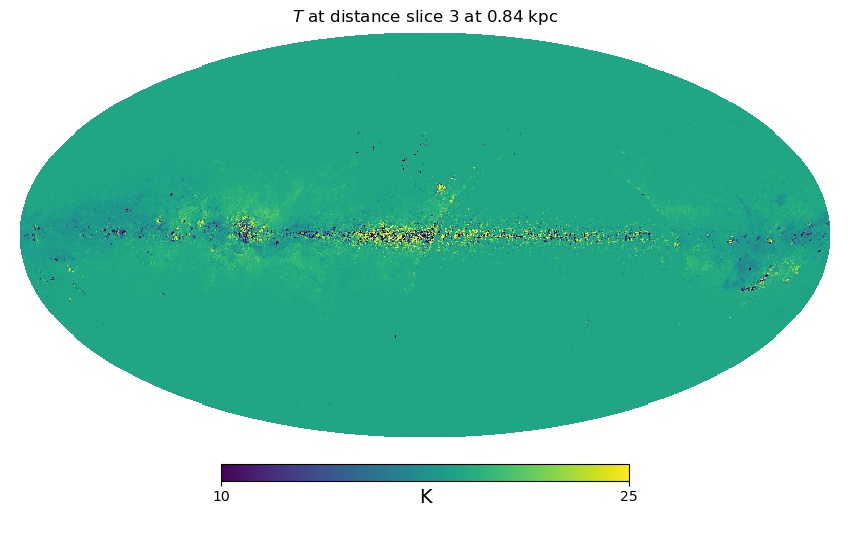} \\
	\end{tabular}
	\vspace{-2.00mm}
	\caption[Temperature of the ISM in 3D]{ Maps of the temperature of dust at different distance slices. The left columns shows the results from fits performed with superpixel Nside of 64, and the right one with Nside of 128. A Gaussian prior centered at 18K with $\sigma$ of 9K is imposed on the temperature parameters in order to constrain the fit in the voxels where dust is not present. As the maps increase in distance, dust becomes present only at lower and lower latitudes. As a result, the higher latitudes become dominated by the temperature prior.}
	\label{fig:dust_3D_temp_3D}
\end{figure*}

For this application, we run the fits for each superpixel in an Nside 32 map. We split each line of sight into 9 voxels, and perform a fit where the temperature is allowed to vary in each voxel, and $\rho$ and $\beta$ are allowed to take a single value per superpixel. Since we want to explore the entire sky, we use the optimizing method described in Section \S \ref{sec:dust_3D_optimizer}. Figs. \ref{fig:reconstructed_dust_emission_sky_maps}, \ref{fig:rho_and_beta}, and
\ref{fig:dust_3D_temp_3D} show our results. 

Most importantly, we see that the fit successfully reconstructs the 2D emission maps from the 3D reddening maps (Fig. \ref{fig:reconstructed_dust_emission_sky_maps}). There are a couple of areas where the fit does not work, in the directions in the sky where there likely is dust beyond the distance bins measured in Bayestar and used in the fit. 

We now have slices that show the temperature of the dust voxels at various distances along the line of sight, thus creating a 3D temperature map of the ISM dust (Figs. \ref{fig:3D_map}, \ref{fig:dust_3D_temp_3D}). While the distance sampling is now quite coarse, it confirms this method will be a useful tool to apply on the future 3D reddening maps that will be much improved.

\begin{figure*}[th!]
    \begin{center}
    		\includegraphics[scale=1]{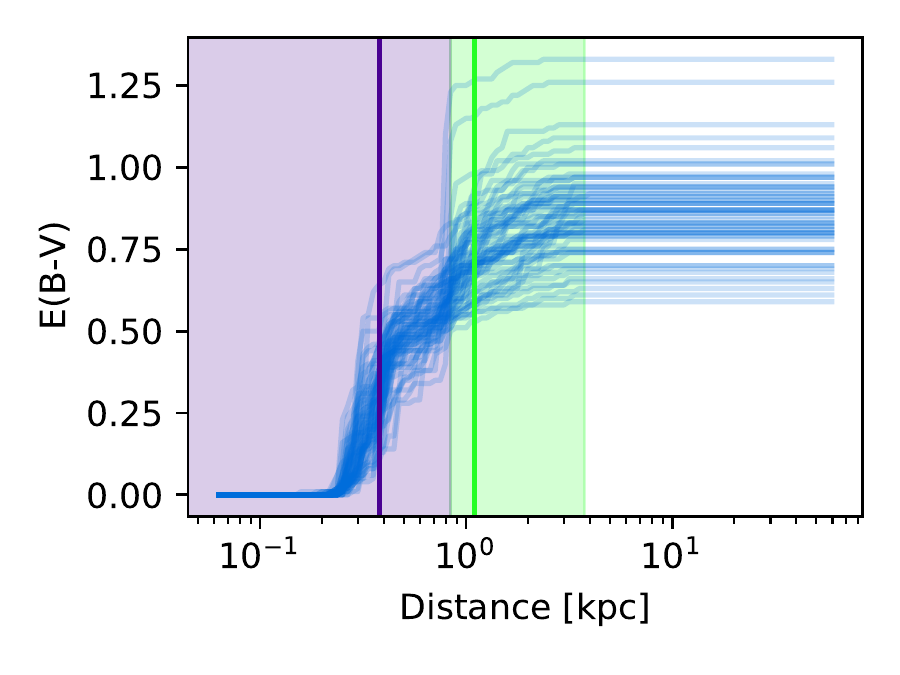} 
    \end{center}
    \vspace{-4.00mm}

	\caption[Cepheus E(B-V) Distance Cut]{The Cepheus dust cloud is known to have two components located at different distances \citep[see Fig. 4]{Zucker2019}. Here we plot the increase in reddening that marks the presence of the two components, in the HEALPix value of 128, centered at l=98.4375$^{\circ}$, b= 8.385$^{\circ}$. The vertical  purple and green lines mark the estimated centers of the two clouds (at 380pc and 1100pc), and the purple and green shaded areas mark the two distance cuts used for this fit.}
	\label{fig:cepheus_EBV_cut}
\end{figure*}

\begin{figure*}[th!]
	\centering
	\begin{tabular}{cccc}
    	\textbf{(a)}  & \textbf{(b)} \\[6pt]
		\hspace{-3.20mm}
		\includegraphics[width=0.50\textwidth]{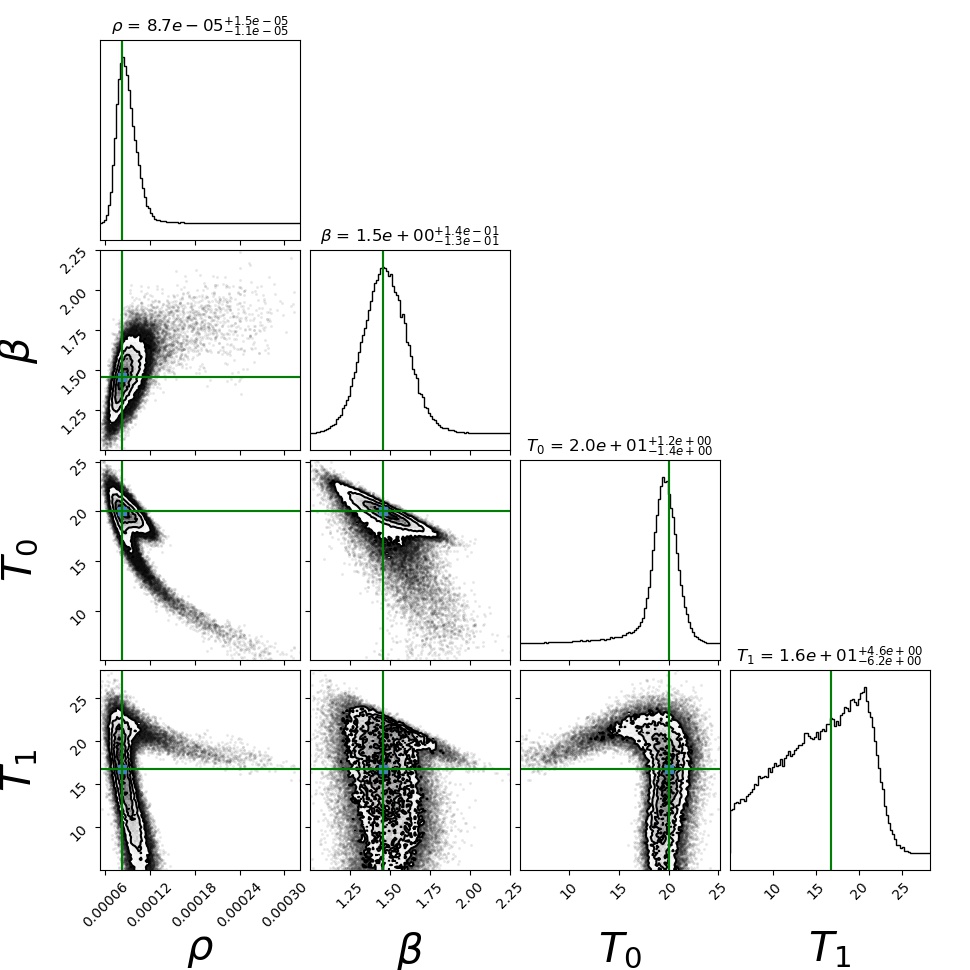} &
		\hspace{-2.70mm}	
		\includegraphics[width=0.50\textwidth]{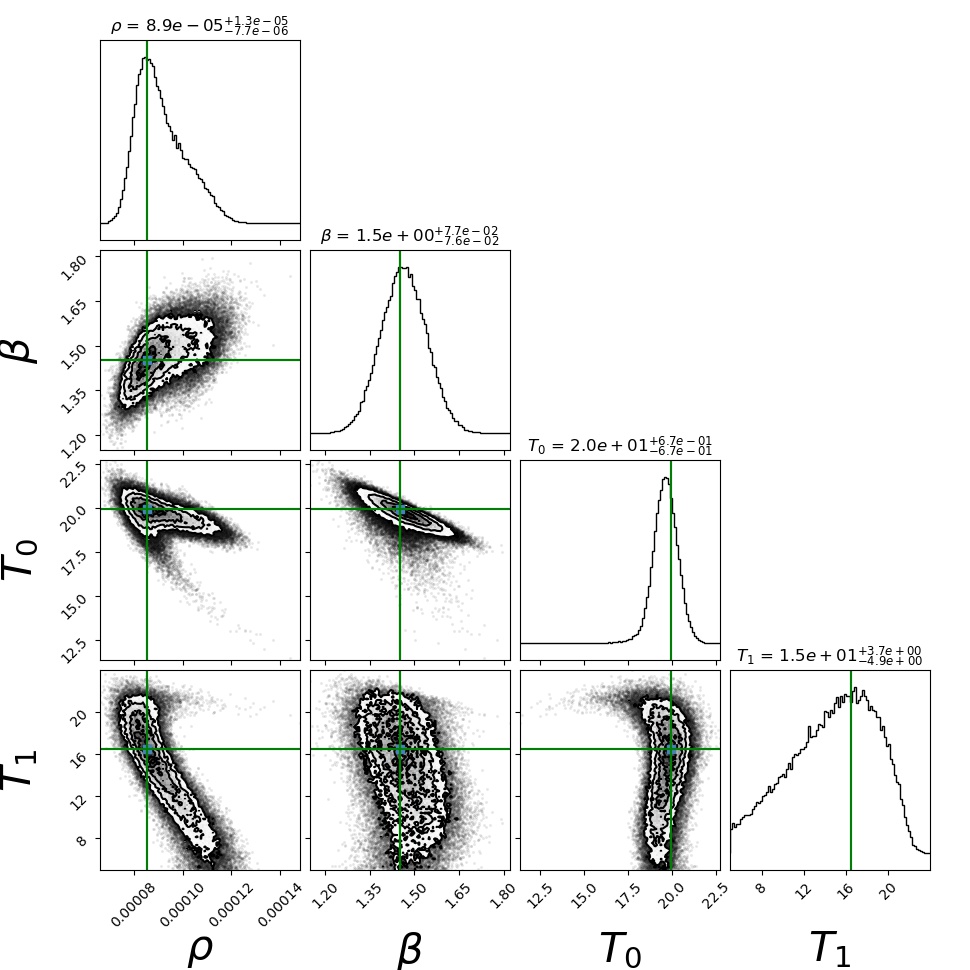} \\
	\end{tabular}
	\vspace{-4.00mm}
	\begin{tabular}{cccc}
    	\textbf{(a)}  & \textbf{(b)} \\[6pt]
		\hspace{-6mm}
		\includegraphics[width=0.50\textwidth]{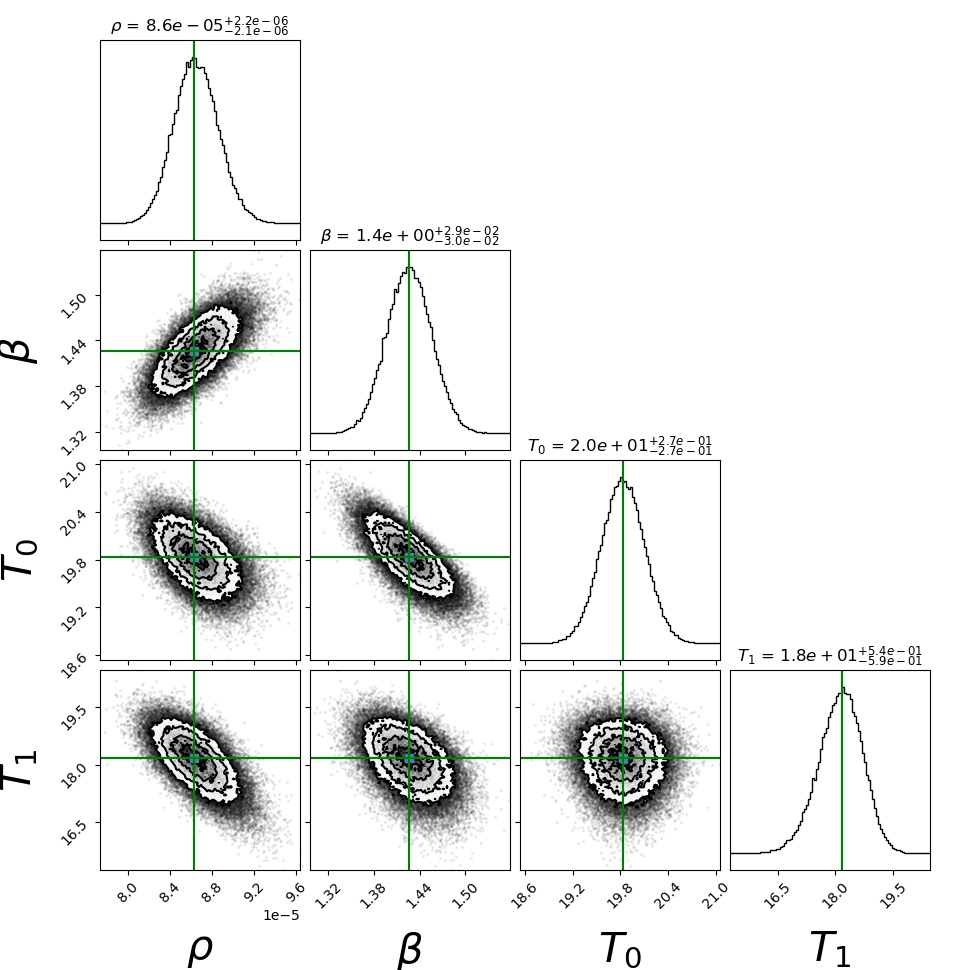} &
		\hspace{-5mm}	
		\includegraphics[width=0.50\textwidth]{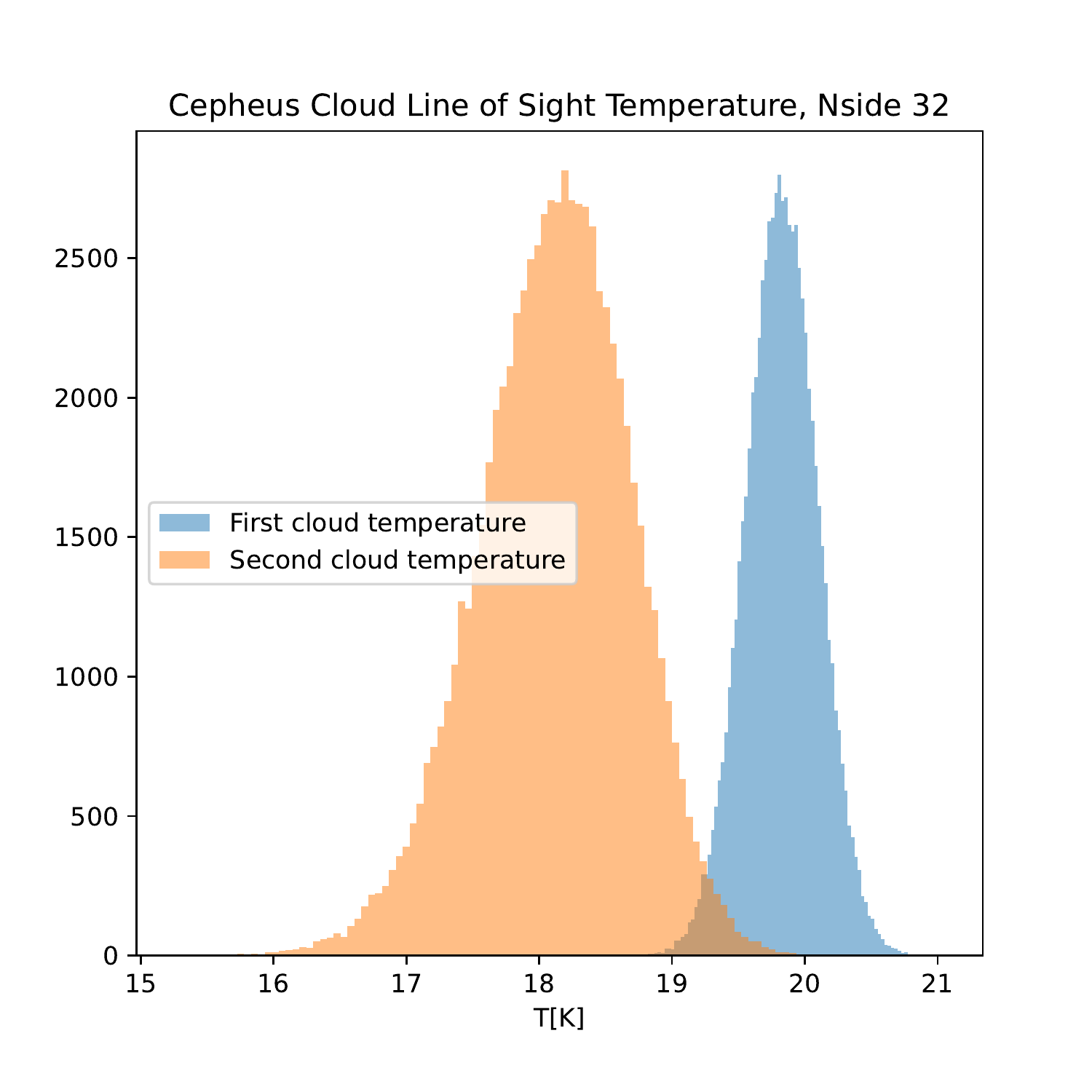} \\
	\end{tabular}
	\caption[Sampler Results for the Cepheus Cloud]{Sampler Results for the Cepheus Cloud. The Nsize of the superpixel took different values for the 3 runs: 128 for (a), 64 for (b), and 32 for (c) and (d). We see the posteriors become much more constrained as the Nside decreases. In the Nside 32 case, we can see a difference between the dust temperatures in the two voxels along the line of sight, indicating that the two clouds/components  have different temperatures. In the first three figures, the green lines represent the value obtained from the optimizer, which was run first, and which provided the parameter space from which to generate the initialization positions for the walkers of the sampler.}
	
	\label{fig:cepheus_sampler}
\end{figure*}
\subsection{3D visualisation of the 3D dust temperature map}


We generate 3D visualizations (Fig. \ref{fig:3D_map}) that show both dust density and temperature. More technically, we use volumetric ray-casting to generate color volume renderings of the dust, in which the overall emissivity of the dust is proportional to its density, with a temperature-dependent term that is different in the red, green and blue (RGB) channels. In a given volume rendering, the RGB value $v_i$ (where $i$ indexes the color channel) of each pixel is calculated by integrating our emissivity function $\varepsilon_i(\vec{r})$ along a ray (parameterized by the normal vector $\hat{n}$) emanating from our virtual camera, located at $\vec{r}_0$:
\begin{align}
    v_i
    =
    \frac{1}{v_{\mathrm{max}}}
    \int_0^{l_{\mathrm{max}}} \!\!\!\!\!\!\!\!\!
        \varepsilon_i \left( \vec{r}_0 + l \, \hat{n} \right) \mathrm{d}l
    \, ,
\end{align}
where $v_{\mathrm{max}}$ is a normalizing factor and $l_{\mathrm{max}}$ is the maximum distance to which we integrate each ray. The emissivity is proportional to the density of the dust, with temperature-dependence given by the Planck function, and with each channel (R, G or B) sensitive to a different frequency, $\nu_i$:
\begin{align}
    \varepsilon_i \left(\vec{r}\right)
    =
    \tau \left(\vec{r}\right) B_{\nu} \left[ T\left(\vec{r}\right), \nu_i \right]
    \, .
\end{align}
We choose the frequency $\nu_i$ to which each band is sensitive by choosing a temperature $T_i$ and converting to frequency using Wien's displacement law. We choose $T = \left( 16, 18, 20 \right) \,\mathrm{K}$ for the R, G and B channels, respectively. After ray-casting the entire image, we clip the pixel values to the range $\left[0, 1\right]$ and perform a gamma stretch (with $\gamma = 0.25$) of the saturation of the images (in hue-saturation-value-space), in order to make temperature variations in the map more easily visible to the eye.

\subsection{Variation of the $\rho$ conversion factor}

Emission-based interstellar dust maps, e.g.\ \cite{Schlegel1998}, have been a valuable tool for predicting extinction across the sky.
They make the assumption that the ratio of near-infrared extinction to the emission optical depth does not vary with $R_{\textnormal{V}}$, or other dust parameters, across the sky.

Our fit allowed $\rho$ to vary for each superpixel, which resulted in an Nside 32 map of the $\rho$ (see Fig. \ref{fig:rho_and_beta} panel a). The distribution of the resulting $\rho$ values gives $\rho=1.1e-04_{-3.0e-05}^{ +2.5e-04}$. Thus, we find $\rho$ to have variable values across the sky, and this variation may be meaningful depending on the application.

This result also connects to Fig.\ 22 of \cite{Zelko2020}, which found that certain dust models lead to a dependence of $\rho$ on $R_{\textnormal{V}}$

\subsection{3D Temperature of the Cepheus cloud}\label{sec:cepheus_cloud_temperature}

In addition to obtaining the sky temperature maps using an optimizer, we would like to know what precision we have for the temperature measurement. To achieve this purpose, we focus on mapping the temperature of a near-by dust cloud with two components, Cepheus \citep[see Fig. 4]{Zucker2019}. We pick a line of sight in the Cepheus cloud that has distinct contribution from the two components, and set the distance bin cuts for the E(B-V) to contain each of them (Fig. \ref{fig:cepheus_EBV_cut}).

We use the sampler described in Section \S \ref{sec:dust_3D_sampling} to run the analyses. 3 different models are used, where the superpixel Nside takes the values 128, 64, 32 (Fig. \ref{fig:cepheus_sampler}). $\beta$ and $\rho$ are kept constant along the line of sight, but are allowed to vary for each individual superpixel. The results show that as the Nside of the superpixel increases, the spread in the posterior of the parameter values becomes smaller and smaller. This is most likely a result of the fit having better access to the structure of the cloud in the data at each distance, as the superpixel size increases. Finally, as the Nside of the superpixel reaches 32, the posteriors become very close to Gaussian, and, in the case for this line of sight and these distance slices, also show different temperatures, outside of the 1-$\sigma$ error bar. Thus, while we do not know if the model is accurate, there is enough constraint power for it to be precise.

\section{Applications of the 3D dust temperature map}\label{sec:3D_dust_applications}



The 3D dust temperature analysis is limited by the quality of the 3D reddening map. At the current state, our analysis's main intent is proof of concept of the method, which given future improvements in reddening maps, has the potential to achieve many goals. A 3D dust temperature map and emission model would serve as a building block for follow up for studies of the ISM, the magnetic field of our galaxy, and cosmological studies. This section describes these possible future applications. 

\paragraph{First 3D map of the galactic dust temperature and beyond}
A comprehensive 3D map of the properties of dust can serve as a great probe for the interstellar medium, as well as the radiation field, which is scattered up to gamma-ray energies by high-energy electron cosmic rays. 

Building on this 3D dust temperature map, we can aim for two important lines of research: characterizing the 3D structure of the magnetic field, and computing the six-dimensional phase-space density of the interstellar radiation field (the amount of light in every 3D voxel, in every direction, at every energy). The latter can be obtained by using the 3D star distribution and solving the equations of radiative transfer, and computing the dust heating and dust emission in the far infrared. 
Obtaining this data would allow us to see the Galaxy from any vantage point, at any angle, in any color, though with a precision that decreases with increasing distance from the Sun. This would provide a critical element in the model of the diffuse Galactic gamma-ray emission, because inverse-Compton scattering is a significant for source of Galactic gamma-ray emission.  This is especially important for searches for dark matter annihilation. And of course, the biggest payoff from such an effort may be the discovery of the unexpected.

\paragraph{Testing Magnetic Field Models and Tomography of the Magnetic Field}
For the 3D magnetic field, studies have been performed for reduced patches of the sky \citep{Clark2019}, but not yet one on a large scale. There are two possible approaches. First, we can determine the available magnetic field models that would match emission from the 3D dust temperature map with the observed polarization field from the \emph{Planck} satellite. The second approach relies on the fact that EoR experiments accurately measure the polarization of synchrotron at $\sim100$ MHz.  Synchrotron is linearly polarized when it is emitted, based on the magnetic field at that location. A "rotation measure" related to the integral of the electron density and line-of-sight magnetic field can be measured using the frequency dependence of Faraday rotation of the polarization. Using rotation measure synthesis (\cite{Schnitzeler2015}), we will know how much the synchrotron polarization has been rotated along the line of sight, and we can bring in the 3D dust map and constraints from polarization of dust emission to obtain a tomography of the B field.

\paragraph{Studying the relation between $R_{\textrm{V}}$ and $\beta$}
In addition, by doing the 3D dust analysis in the galaxy plane we can obtain spatial correlations of $\tau$, $\beta$, $T$. This can be used as prior information in the analysis for dust at higher latitudes. Thus, analysis for missions like \emph{PIXIE} can be informed about what correlations we can expect in dust, and marginalize over it. Also, right now the error bars are too big when we allow beta to float in each distance bin. However, with future generations of 3D reddening maps, we could test the 2D $R_{\textrm{V}}-\beta$ relation which was the subject of \cite{Zelko2020} and \cite{Schlafly2016} in 3D.

Another potential application is to determine line of sight frequency decorrelation, as discussed in  \cite{Pelgrims2021}. In Section \S \ref{sec:cepheus_cloud_temperature}, we showed that the temperature of different components of a cloud can be determined with enough precision, given a model. This information, combined with more information on magnetic fields, can tell us if the polarization coming from dust has a frequency dependence, and thus if maps of dust measured at high frequencies where dust dominates (e.g. \emph{Planck} 353 GHz) cannot be used to subtract the dust polarization component at other frequencies.




\section{Conclusion}\label{sec:dust_3D_conclussion}
In this work we demonstrate the proof-of-concept for a 3D dust temperature in the interstellar medium, created from the existing 3D maps of dust reddening, and 2D maps of the dust emission. 

To create the map, we cut distance slices along the line of sight,  propose a modified black body emissivity for each of the voxels, and compare the integral along the line of sight with emission observed by \emph{Planck} and \emph{IRAS} at five frequency bands, over the area of a superpixel of Nside smaller than the data maps. Thus, we can perform a large statistical study of the 3D temperature of the dust in the galaxy. 

The results show that emission map of dust can be reconstructed very well from the 3D dust reddening map derived from stellar light absorption, and temperature values can be obtained for distance slices of choice. 

The temperature for a two-component dust cloud test, Cepheus, is explored with a Bayesian sampler; the results show a distinct temperature for the two components of the cloud located at different distances. This would be an important result for dust frequency decorrelation analysis which would be impacted by a line-of-sight with different temperature and magnetic field components. 

Finally, this work also explores the variation across the sky of the conversation factor between emission dust optical depth and reddening. This conversion factor is assumed to be constant in emission-based reddening maps (SFD, \emph{Planck}). The fact that it appears to vary suggests that this assumption should be revisited, possibly leading to a recalibration of commonly used reddening maps.


In the future, datasets like \emph{LSST} \cite{Collaboration2009} will drastically improve 3D reddening maps. While our fits may not enough precision for certain applications at the current stage, once the future 3D reddening maps come out, the quality in the 3D dust temperature fits will also increase significantly.
\paragraph{Acknowledgments}
We acknowledge helpful conversations with  Susan Clark, Cora Dvorkin,  Daniel Eisenstein, Brandon Hensley, John Kovac, and Catherine Zucker. \\

IZ was supported by the Harvard College Observatory. DF is partially supported by  NSF grant AST-1614941, ``Exploring the Galaxy: 3-Dimensional Structure and Stellar Streams.'' 
This research made use of the NASA Astrophysics Data System Bibliographic Services (ADS), the Odyssey Cluster at Harvard University, the color blindness palette by Martin Krzywinski \& Jonathan Corum\footnote{\url{http://mkweb.bcgsc.ca/biovis2012/color-blindness-palette.png}}, and the Color Vision Deficiency PDF Viewer by Marrie Chatfield \footnote{\url{https://mariechatfield.com/simple-pdf-viewer/}}.

\paragraph{software} healpy \citep{Gorski2005, Zonca2019}, ptemcee \citep{Vousden2016}, NumPy \citep{VanderWalt2011}, Matplotlib \citep{Hunter2007}, pandas \citep{mckinney-proc-scipy-2010}, scikit-learn \citep{Pedregosa2012}, IPython \citep{Perez2007}, Python \citep{Millman2011, Oliphant2007}, dustmaps \citep{Green2018a}, Mathematica\citep{Mathematica}

\appendix

\section{Optimizer}
\label{sec:appendix}

The model described in Section \S \ref{sec:dust_3D_model} allows us to calculate the gradient and the Hessian matrix for all the parameters of the fit. This can come in handy, as a function implementation giving the gradient and Hessian matrix can be passed to an optimizer or sampling method, reducing the computation time by a lot, since the method of choice no longer needs to estimate these quantities numerically.

\subsection{Gradient}
The function that we want to obtain the gradient of is given by equation \ref{eq:chi_square_2}. In this case, it also includes Gaussian priors on $T, \beta, $ and $\rho$.

\begin{equation}{\label{eq:chi_square_2}}
\begin{split}
f &= \Delta \chi^2 + \sum_{n}\left(\frac{(T^n-T_0)^2}{{\sigma_{T}}^2} +
                     \frac{(\beta^n-\beta_0)^2}{{\sigma_{\beta}}^2} +
                     \frac{(\rho_{353}^{n}-\rho_0)^2}{{\sigma_{\rho}}^2}
                     \right)\\
  &= \sum_{k,\nu} \left (\left ( O_{\nu} + \sum_n \Delta E_{\textrm{B-V}}^{k,n} \rho_{353}^n  \big ( \frac{\nu}{\nu_0} \big )^{\beta^n} B_{\nu}(T^n) - I_{\nu.k}^{\textrm{D}}\right)/\sigma_{\nu}^{\textrm{D}}\right)^2 + \\
  &\;\;\;\; + 
  \sum_{n}\left(\frac{(T^n-T_0)^2}{{\sigma_{T}}^2} +
                     \frac{(\beta^n-\beta_0)^2}{{\sigma_{\beta}}^2} +
                     \frac{(\rho_{353}^{n}-\rho_0)^2}{{\sigma_{\rho}}^2}
                     \right)
\end{split}
\end{equation}

To make it easier to keep track of the differentiation process, we define the following helper functions:
\paragraph{Helper Functions}
\begin{equation}{\label{eq:g}}
g_{k,\nu} = 2\left ( O_{\nu} + \sum_n  \Delta E_{\textrm{B-V}}^{k,n} \rho_{353}^n \big ( \frac{\nu}{\nu_0} \big )^{\beta^n} B_{\nu}(T^n) - I_{\nu,k}^{\textrm{D}}\right)/{\sigma_{\nu}^{\textrm{D}}}^2
\end{equation}

\begin{equation}
    A_{\nu}^{k,n} =  \Delta E_{\textrm{B-V}}^{k,n}      \big ( \frac{\nu}{\nu_0} \big )^{\beta^n} B_{\nu}(T^n)
\end{equation}

\begin{equation}
    B_{\nu}^{k,n} =  \Delta E_{\textrm{B-V}}^{k,n}  \rho^n_{353} \big ( \frac{\nu}{\nu_0} \big )^{\beta^n} B_{\nu}(T^n) \ln{\frac{\nu}{\nu_0}}
\end{equation}
\begin{equation}
    C_{\nu}^{k,n} = \Delta E_{\textrm{B-V}}^{k,n} \rho_{353}^n  \big ( \frac{\nu}{\nu_0} \big )^{\beta^n} \dv{B_{\nu}(T^n)}{T^n}
\end{equation}
\begin{equation}
    D_{\nu}^{k,n} = \Delta E_{\textrm{B-V}}^{k,n}   \big ( \frac{\nu}{\nu_0} \big )^{\beta^n} \dv{B_{\nu}(T^n)}{T^n}
\end{equation}
\begin{equation}
    F_{\nu}^{k,n} = \Delta E_{\textrm{B-V}}^{k,n} \rho_{353}^n  \big ( \frac{\nu}{\nu_0} \big )^{\beta^n} \dv[2]{B_{\nu}(T^n)}{{T^n}}
\end{equation}

Thus, we now obtain the gradient of $f$, with its elements given by:
\paragraph{Gradient}
\begin{equation}\label{eq:gradient_O}
\pdv{f}{O_{\nu}} = \sum_{k} g_{k,\nu} 
\end{equation}

\begin{equation}\label{eq:gradient_rho}
\pdv{f}{\rho^n_{353}} = \sum_{k,\nu} g_{k,\nu} \Delta E_{\textrm{B-V}}^{k,n} \big ( \frac{\nu}{353\text{  GHz}} \big )^{\beta^n}                         B_{\nu}(T^n)+ \frac{2(\rho_{353}^{n}-\rho_0)}{{\sigma_{\rho}}^2}
                      = \sum_{k,\nu} g_{k,\nu} A_{\nu}^{k,n} + \frac{2(\rho_{353}^{n}-\rho_0)}{{\sigma_{\rho}}^2}
\end{equation}

if $\rho$ is fixed along the sightline, this becomes 
\begin{equation}\label{eq:gradient_rho_fixed}
\pdv{f }{\rho_{353}} =  \sum_{k,\nu} g_{k,\nu} \sum_n   \Delta E_{\textrm{B-V}}^{k,n} \big ( \frac{\nu}{\nu_0} \big )^{\beta^n} B_{\nu}(T^n)+
\frac{2(\rho_{353}-\rho_0)}{{\sigma_{\rho}}^2}
                     = \sum_{k,\nu} g_{k,\nu} \sum_n A_{\nu}^{k,n} + \frac{2(\rho_{353}-\rho_0)}{{\sigma_{\rho}}^2}
\end{equation}

\begin{equation}\label{eq:gradient_beta}
\pdv{f }{\beta^n} =   \sum_{k,\nu} g_{k,\nu}\Delta E_{\textrm{B-V}}^{k,n} \rho_{353}^n  \big ( \frac{\nu}{\nu_0} \big )^{\beta^n}B_{\nu}(T^n) \ln{\frac{\nu}{\nu_0}}+ \frac{2(\beta^n-\beta_0)}{{\sigma_{\beta}}^2}
                  = \sum_{k,\nu} g_{k,\nu} B_{\nu}^{k,n} + \frac{2(\beta^n-\beta_0)}{{\sigma_{\beta}}^2}
\end{equation}

if $\beta$ is fixed along the sightline, this becomes 

\begin{equation}\label{eq:gradient_beta_fixed}
\pdv{f }{\beta} = \sum_{k,\nu} g_{k,\nu}\sum_n \Delta E_{\textrm{B-V}}^{k,n} \rho_{353}^n  \big ( \frac{\nu}{\nu_0} \big )^{\beta}B_{\nu}(T^n) \ln{\frac{\nu}{\nu_0}}+\frac{2(\beta-\beta_0)}{{\sigma_{\beta}}^2} 
                  = \sum_{k,\nu} g_{k,\nu} \sum_n B_{\nu}^{k,n} +\frac{2(\beta-\beta_0)}{{\sigma_{\beta}}^2} 
\end{equation}

\begin{equation}\label{eq:gradient_T}
\pdv{f }{T^n} =  \sum_{k,\nu} g_{k,\nu} \Delta E_{\textrm{B-V}}^{k,n} \rho_{353}^n  \big ( \frac{\nu}{\nu_0} \big )^{\beta^n} \dv{B_{\nu}(T^n)}{T^n}+ \frac{2(T^n-T_0)}{{\sigma_{T}}^2} = \sum_{k,\nu} g_{k,\nu} C_{\nu}^{k,n} + \frac{2(T^n-T_0)}{{\sigma_{T}}^2}
\end{equation}

if $T$ is fixed along the sightline, this becomes
\begin{equation}\label{eq:gradient_T_fixed}
\pdv{f}{T} =  \sum_{k,\nu} g_{k,\nu} \sum_n \Delta E_{\textrm{B-V}}^{k,n} \rho_{353}^n  \big ( \frac{\nu}{\nu_0} \big )^{\beta^n} \dv{B_{\nu}(T)}{T}+ \frac{2(T-T_0)}{{\sigma_{T}}^2} = \sum_{k,\nu} g_{k,\nu}\sum_n  C_{\nu}^{k,n} + \frac{2(T-T_0)}{{\sigma_{T}}^2}
\end{equation}

\subsection{Hessian Matrix}
We calculate the Hessian matrix for the function \ref{eq:chi_square_2}. To do so, we first take the derivatives of the helper functions already defined:
\subsubsection{Helper functions derivatives}
\begin{equation}
    \pdv{g_{k,\nu_i}}{O_{\nu_j}} = \delta_{ij} \frac{2}{{\sigma_{\nu}^{\textrm{D}}}^2};\;
    \pdv{g_{k,\nu}}{\rho_{353}^{n}} = \frac{2}{{\sigma_{\nu}^{\textrm{D}}}^2} \Delta E_{\textrm{B-V}}^{k,n}      \big ( \frac{\nu}{\nu_0} \big )^{\beta^n} B_{\nu}(T^n)=\frac{2}{{\sigma_{\nu}^{\textrm{D}}}^2} A_{\nu}^{k,n} \;
\end{equation}
\begin{equation}
     \pdv{g_{k,\nu}}{\rho_{353}} = \frac{2}{{\sigma_{\nu}^{\textrm{D}}}^2} \sum_n \Delta E_{\textrm{B-V}}^{k,n}     \big ( \frac{\nu}{\nu_0} \big )^{\beta^n} B_{\nu}(T^n) = \frac{2}{{\sigma_{\nu}^{\textrm{D}}}^2} \sum_n A_{\nu}^{k,n} 
\end{equation}
\begin{equation}
    \pdv{g_{k,\nu}}{\beta^n} =  \frac{2}{{\sigma_{\nu}^{\textrm{D}}}^2} \Delta E_{\textrm{B-V}}^{k,n}          \rho_{353}^n  \big ( \frac{\nu}{\nu_0} \big )^{\beta^n}B_{\nu}(T^n)
        \ln{\frac{\nu}{\nu_0}} =\frac{2}{{\sigma_{\nu}^{\textrm{D}}}^2} B_{\nu}^{k,n} \;
\end{equation}
\begin{equation}
    \pdv{g_{k,\nu}}{\beta} =  \frac{2}{{\sigma_{\nu}^{\textrm{D}}}^2} \sum_n \Delta E_{\textrm{B-V}}^{k,n}          \rho_{353}^n  \big ( \frac{\nu}{\nu_0} \big )^{\beta^n}B_{\nu}(T^n)
        \ln{\frac{\nu}{\nu_0}}   = \frac{2}{{\sigma_{\nu}^{\textrm{D}}}^2} \sum_n B_{\nu}^{k,n}
\end{equation}

\begin{equation}
    \pdv{g_{k,\nu}}{T^n} =  \frac{2}{{\sigma_{\nu}^{\textrm{D}}}^2} \Delta E_{\textrm{B-V}}^{k,n} \rho_{353}^n  \big ( \frac{\nu}{\nu_0} \big )^{\beta^n} \dv{B_{\nu}(T^n)}{T^n} = \frac{2}{{\sigma_{\nu}^{\textrm{D}}}^2} C_{\nu}^{k,n} 
\end{equation}

Thus, we can now obtain the elements of the Hessian matrix (also shown in Table \ref{table:hessian_matrix}):

\begin{equation}
    \pdv{f}{O_{\nu_i}}{O_{\nu_j}} = \sum_k\delta_{ij} \frac{2}{{\sigma_{\nu}^{\textrm{D}}}^2} = \delta_{i,j}\frac{2M}{{\sigma_{\nu}^{\textrm{D}}}^2}
\end{equation}
where $M$ is the number of pixels in a superpixel.
\begin{equation}
    \pdv{f}{O_{\nu}}{\rho^n_{353}} = \sum_k \frac{2}{{\sigma_{\nu}^{\textrm{D}}}^2} A_{\nu}^{k,n};\; 
    \pdv{f}{\rho^n_{353}}{O_{\nu_j}} = \sum_{k,\nu_i} \pdv{g_{k,\nu_i}}{O_{\nu_j}} A_{\nu_i}^{k,n} = \sum_{k,\nu_i} \delta_{i,j} \frac{2}{{\sigma_{\nu_j}^{\textrm{D}}}^2} A_{\nu_i}^{k,n} = \sum_{k} \frac{2}{{\sigma_{\nu_j}^{\textrm{D}}}^2} A_{\nu_j}^{k,n}
\end{equation}
$\rho$
\begin{equation}
    \pdv{f}{\rho^{n_i}_{353}}{\rho^{n_j}_{353}} = \sum_{k,\nu} \pdv{g_{k,\nu}}{\rho^{n_j}_{353}} A_{\nu}^{k,n_i} + \delta_{ij}\frac{2}{{\sigma_{\rho}}^2} = \sum_{k,\nu} \frac{2}{{\sigma_{\nu}^{\textrm{D}}}^2} A_{\nu}^{k,n_j} A_{\nu}^{k,n_i} + \delta_{ij}\frac{2}{{\sigma_{\rho}}^2}
\end{equation}

\begin{equation}
    \pdv{f}{\rho^{n_i}_{353}}{\beta^{n_j}} =  \sum_{k,\nu} \pdv{g_{k,\nu}}{\beta^{n_j}} A_{\nu}^{k,n_i} + \sum_{k,\nu} g_{k,\nu} \pdv{A_{\nu}^{k,n_i}}{\beta^{n_j}} = \sum_{k,\nu}\left( \frac{2}{{\sigma_{\nu}^{\textrm{D}}}^2} B_{\nu}^{k,n_j} A_{\nu}^{k,n_i} + \delta_{ij} g_{k,\nu} A_{\nu}^{k,n_i}\ln{\frac{\nu}{\nu_0}} \right)
\end{equation}
\begin{equation}
\pdv{f}{\rho^{n_i}_{353}}{T^{n_j}} = \sum_{k,\nu} \left(\frac{2}{{\sigma_{\nu}^{\textrm{D}}}^2} C_{\nu}^{k,n_j} A_{\nu}^{k,n_i}+\delta_{ij} g_{k,\nu}D_{\nu}^{k,n_i}\right)
\end{equation}
$\beta$
\begin{equation}
\begin{split}
    \pdv{f}{\beta^{n_i}}{\beta^{n_j}}= &\sum_{k,\nu} \pdv{g_{k,\nu}}{\beta^{n_j}} B_{\nu}^{k,n_i} + \sum_{k,\nu} g_{k,\nu} \pdv{B_{\nu}^{k,n_i}}{\beta^{n_j}}+ \delta_{ij}\frac{2}{{\sigma_{\beta}}^2} \\
    &=  \sum_{k,\nu} \left( \frac{2}{{\sigma_{\nu}^{\textrm{D}}}^2} B_{\nu}^{k,n_j} B_{\nu}^{k,n_i} + \delta_{ij} g_{k,\nu}B_{\nu}^{k,n_i}\ln{\frac{\nu}{\nu_0}}\right) + \delta_{ij}\frac{2}{{\sigma_{\beta}}^2}
\end{split}
\end{equation}
$T$
\begin{equation}
\pdv{f}{T^{n_i}}{\beta^{n_j}} =   \sum_{k,\nu} \pdv{g_{k,\nu}}{\beta^{n_j}} C_{\nu}^{k,n_i} + \sum_{k,\nu} g_{k,\nu} \pdv{C_{\nu}^{k,n_i}}{\beta^{n_j}} = \sum_{k,\nu}\left( \frac{2}{{\sigma_{\nu}^{\textrm{D}}}^2} B_{\nu}^{k,n_j} C_{\nu}^{k,n_i} + \delta_{ij} g_{k,\nu} C_{\nu}^{k,n_i}\ln{\frac{\nu}{\nu_0}} \right)
\end{equation}

\begin{equation}
\pdv{f}{T^{n_i}}{T^{n_j}} =   \sum_{k,\nu} \pdv{g_{k,\nu}}{T^{n_j}} C_{\nu}^{k,n_i} + \sum_{k,\nu} g_{k,\nu} \pdv{C_{\nu}^{k,n_i}}{T^{n_j}} = \sum_{k,\nu}\left( \frac{2}{{\sigma_{\nu}^{\textrm{D}}}^2} C_{\nu}^{k,n_j} C_{\nu}^{k,n_i} + \delta_{ij} g_{k,\nu} F_{\nu}^{k,n_i} \right) + \delta_{ij}\frac{2}{{\sigma_{T}}^2}
\end{equation}

\begin{sidewaystable}

\vspace{2cm}
\begin{center}
\begin{tabular}{ |c|c|c|c|c| } 
 \hline
          &             $O_{\nu}$            &             $\rho^n$                          &            $\beta^n$                     &             $T^n$                    \\[1mm] 
$O_{\nu}$ & $\pdv{f}{O_{\nu_i}}{O_{\nu_j}}$  & $\pdv{f}{O_{\nu}}{\rho^n_{353}}$              & $\pdv{f}{O_{\nu}}{\beta^n}$              & $\pdv{f}{O_{\nu}}{T^n}$              \\[2.5mm]
$\rho^n$  & $\pdv{f}{\rho^n_{353}}{O_{\nu}}$ & $\pdv{f}{\rho^{n_i}_{353}}{\rho^{n_j}_{353}}$ & $\pdv{f}{\rho^{n_i}_{353}}{\beta^{n_j}}$ & $\pdv{f}{\rho^{n_i}_{353}}{T^{n_j}}$ \\[2.5mm]
$\beta^n$ & $\pdv{f}{\beta^n}{O_{\nu}}$      & $\pdv{f}{\beta^{n_i}}{\rho^{n_j}_{353}}$      & $\pdv{f}{\beta^{n_i}}{\beta^{n_j}}$      & $\pdv{f}{\beta^{n_i}}{T^{n_j}}$      \\[2.5mm]
$T^n$     & $\pdv{f}{T^n}{O_{\nu}}$          & $\pdv{f}{T^{n_i}}{\rho^{n_j}_{353}}$         & $\pdv{f}{T^{n_i}}{\beta^{n_j}}$          & $\pdv{f}{T^{n_i}}{T^{n_j}}$          \\[2.5mm]
 \hline
\end{tabular}
\end{center}

\hspace{-1cm}
\begin{tabular}{ |c|c|c|c|c| } 
 \hline
          &             $O_{\nu}$            &             $\rho^n$                          &            $\beta^n$                     &             $T^n$                    \\[1mm] 
$O_{\nu}$ &  $\delta_{ij}\frac{2M}{{\sigma_{\nu}^{\textrm{D}}}^2}$  & $\sum_k \frac{2}{{\sigma_{\nu}^{\textrm{D}}}^2} A_{\nu}^{k,n}$  & $\sum_k \frac{2}{{\sigma_{\nu}^{\textrm{D}}}^2} B_{\nu}^{k,n}$  & $\sum_k \frac{2}{{\sigma_{\nu}^{\textrm{D}}}^2} C_{\nu}^{k,n}$ \\[2.5mm]
$\rho^n$  &$\sum_k \frac{2}{{\sigma_{\nu}^{\textrm{D}}}^2} A_{\nu}^{k,n}$ 
&$\sum_{k,\nu} \frac{2}{{\sigma_{\nu}^{\textrm{D}}}^2} A_{\nu}^{k,n_j} A_{\nu}^{k,n_i} + \delta_{ij}\frac{2}{{\sigma_{\rho}}^2}$ 
&$\sum_{k,\nu}\left( \frac{2}{{\sigma_{\nu}^{\textrm{D}}}^2} B_{\nu}^{k,n_j} A_{\nu}^{k,n_i} + \delta_{ij} g_{k,\nu} A_{\nu}^{k,n_i}\ln{\frac{\nu}{\nu_0}} \right)$
& $\sum_{k,\nu} \left(\frac{2}{{\sigma_{\nu}^{\textrm{D}}}^2} C_{\nu}^{k,n_j} A_{\nu}^{k,n_i}+\delta_{ij} g_{k,\nu}D_{\nu}^{k,n_i}\right)$ \\[2.5mm]
$\beta^n$ & $\sum_k \frac{2}{{\sigma_{\nu}^{\textrm{D}}}^2} B_{\nu}^{k,n}$   
& $\sum_{k,\nu}\left( \frac{2}{{\sigma_{\nu}^{\textrm{D}}}^2}A_{\nu}^{k,n_i} B_{\nu}^{k,n_i}  + \delta_{ij} g_{k,\nu} A_{\nu}^{k,n_i}\ln{\frac{\nu}{\nu_0}} \right)$      
& $\sum_{k,\nu} \left( \frac{2}{{\sigma_{\nu}^{\textrm{D}}}^2} B_{\nu}^{k,n_j} B_{\nu}^{k,n_i} + \delta_{ij} g_{k,\nu}B_{\nu}^{k,n_i}\ln{\frac{\nu}{\nu_0}}\right) + \delta_{ij}\frac{2}{{\sigma_{\beta}}^2}$      
& $\sum_{k,\nu}\left( \frac{2}{{\sigma_{\nu}^{\textrm{D}}}^2}C_{\nu}^{k,n_j}  B_{\nu}^{k,n_i} + \delta_{ij} g_{k,\nu} C_{\nu}^{k,n_i}\ln{\frac{\nu}{\nu_0}} \right)$      \\[2.5mm]
$T^n$     &  $\sum_k \frac{2}{{\sigma_{\nu}^{\textrm{D}}}^2} C_{\nu}^{k,n}$         
& $ \sum_{k,\nu} \left(\frac{2}{{\sigma_{\nu}^{\textrm{D}}}^2} A_{\nu}^{k,n_j} C_{\nu}^{k,n_i} +\delta_{ij} g_{k,\nu}D_{\nu}^{k,n_i}\right)$        
& $\sum_{k,\nu}\left( \frac{2}{{\sigma_{\nu}^{\textrm{D}}}^2} B_{\nu}^{k,n_j} C_{\nu}^{k,n_i} + \delta_{ij} g_{k,\nu} C_{\nu}^{k,n_i}\ln{\frac{\nu}{\nu_0}} \right)$        
& $\sum_{k,\nu}\left( \frac{2}{{\sigma_{\nu}^{\textrm{D}}}^2} C_{\nu}^{k,n_j} C_{\nu}^{k,n_i} + \delta_{ij} g_{k,\nu} F_{\nu}^{k,n_i} \right) + \delta_{ij}\frac{2}{{\sigma_{T}}^2}$          \\[2.5mm]
 \hline
\end{tabular}


\caption[Hessian Matrix]{The Hessian matrix for the function \ref{eq:chi_square_2}, describing the 3D dust reddening-2D emission fit model used in this work. \label{table:hessian_matrix}}

\end{sidewaystable}

\bibliography{dust_3d_temp.bib}
\end{document}